\PassOptionsToPackage{unicode}{hyperref}
\PassOptionsToPackage{hyphens}{url}
\PassOptionsToPackage{dvipsnames,svgnames,x11names}{xcolor}
\documentclass[12pt]{article}

\usepackage[autonum,colorhypersetup]{shortex}
\usepackage{amsmath,amsfonts,amssymb,amsthm}
\usepackage{enumitem}
\usepackage{iftex}
\ifPDFTeX
  \usepackage[T1]{fontenc}
  \usepackage[utf8]{inputenc}
  \usepackage{textcomp} 
\else 
  \usepackage{unicode-math}
  \defaultfontfeatures{Scale=MatchLowercase}
  \defaultfontfeatures[\rmfamily]{Ligatures=TeX,Scale=1}
\fi
\usepackage{lmodern}
\ifPDFTeX\else  
\fi
\IfFileExists{upquote.sty}{\usepackage{upquote}}{}
\IfFileExists{microtype.sty}{
  \usepackage[]{microtype}
  \UseMicrotypeSet[protrusion]{basicmath} 
}{}
\makeatletter
\@ifundefined{KOMAClassName}{
  \IfFileExists{parskip.sty}{%
    \usepackage{parskip}
  }{
    \setlength{\parindent}{0pt}
    \setlength{\parskip}{6pt plus 2pt minus 1pt}}
}{
  \KOMAoptions{parskip=half}}
\makeatother
\usepackage{xcolor}
\makeatletter
\ifx\paragraph\undefined\else
  \let\oldparagraph\paragraph
  \renewcommand{\paragraph}{
    \@ifstar
      \xxxParagraphStar
      \xxxParagraphNoStar
  }
  \newcommand{\xxxParagraphStar}[1]{\oldparagraph*{#1}\mbox{}}
  \newcommand{\xxxParagraphNoStar}[1]{\oldparagraph{#1}\mbox{}}
\fi
\ifx\subparagraph\undefined\else
  \let\oldsubparagraph\subparagraph
  \renewcommand{\subparagraph}{
    \@ifstar
      \xxxSubParagraphStar
      \xxxSubParagraphNoStar
  }
  \newcommand{\xxxSubParagraphStar}[1]{\oldsubparagraph*{#1}\mbox{}}
  \newcommand{\xxxSubParagraphNoStar}[1]{\oldsubparagraph{#1}\mbox{}}
\fi
\makeatother

\usepackage{longtable,booktabs,array}
\usepackage{calc} 
\usepackage{etoolbox}
\makeatletter
\patchcmd\longtable{\par}{\if@noskipsec\mbox{}\fi\par}{}{}
\makeatother
\IfFileExists{footnotehyper.sty}{\usepackage{footnotehyper}}{\usepackage{footnote}}
\makesavenoteenv{longtable}
\usepackage{graphicx}
\makeatletter
\def\maxwidth{\ifdim\Gin@nat@width>\linewidth\linewidth\else\Gin@nat@width\fi}
\def\maxheight{\ifdim\Gin@nat@height>\textheight\textheight\else\Gin@nat@height\fi}
\makeatother
\setkeys{Gin}{width=\maxwidth,height=\maxheight,keepaspectratio}
\makeatletter
\def\fps@figure{htbp}
\makeatother

\makeatletter
\@ifpackageloaded{caption}{}{\usepackage{caption}}
\AtBeginDocument{%
\ifdefined\contentsname
  \renewcommand*\contentsname{Table of contents}
\else
  \newcommand\contentsname{Table of contents}
\fi
\ifdefined\listfigurename
  \renewcommand*\listfigurename{List of Figures}
\else
  \newcommand\listfigurename{List of Figures}
\fi
\ifdefined\listtablename
  \renewcommand*\listtablename{List of Tables}
\else
  \newcommand\listtablename{List of Tables}
\fi
\ifdefined\figurename
  \renewcommand*\figurename{Figure}
\else
  \newcommand\figurename{Figure}
\fi
\ifdefined\tablename
  \renewcommand*\tablename{Table}
\else
  \newcommand\tablename{Table}
\fi
}
\@ifpackageloaded{float}{}{\usepackage{float}}
\floatstyle{ruled}
\@ifundefined{c@chapter}{\newfloat{codelisting}{h}{lop}}{\newfloat{codelisting}{h}{lop}[chapter]}
\floatname{codelisting}{Listing}

\makeatother
\makeatletter
\@ifpackageloaded{caption}{}{\usepackage{caption}}
\@ifpackageloaded{subcaption}{}{\usepackage{subcaption}}
\makeatother

\ifLuaTeX
  \usepackage{selnolig}  
\fi
\usepackage[]{natbib}
\usepackage{bookmark}

\IfFileExists{xurl.sty}{\usepackage{xurl}}{} 
\hypersetup{
  pdftitle={Title},
  pdfauthor={Author 1; Author 2},
  pdfkeywords={3 to 6 keywords, that do not appear in the title},
  colorlinks=true,
  linkcolor={blue},
  filecolor={Maroon},
  citecolor={Blue},
  urlcolor={Blue},
  pdfcreator={LaTeX via pandoc}}

\newcommand{\anon}{1}

\begin{document}

\def\spacingset#1{\renewcommand{\baselinestretch}%
{#1}\small\normalsize} \spacingset{1}


\if1\anon
{
  \title{\bf Extreme Value Inference for CoVaR and Systemic Risk}
  \author{Xiaoting Li\thanks{
    The authors gratefully acknowledge support from the Natural Sciences and Engineering Research Council of Canada.}\hspace{.2cm}\\
    Department of Statistics, University of Manitoba\\
    and \\
    Harry Joe\\
    Department of Statistics, University of British Columbia}
  \maketitle
} \fi

\if0\anon
{
  \bigskip
  \bigskip
  \bigskip
  \begin{center}
    {\LARGE\bf Extreme Value Inference for CoVaR and Systemic Risk}
\end{center}
  \medskip
} \fi

\bigskip
\begin{abstract}
We develop an extreme value framework for CoVaR centered on $v(q \mid p ; C)$, the copula-adjusted probability level, or equivalently, the CoVaR on the 
uniform (0,1) scale. 
We characterize the possible tail regimes of $v(q \mid p ; C)$ through the limit behavior of the copula conditional distribution and show that these regimes are determined by the joint tail expansions of the copula. This leads to tractable conditions for identifying the tail regime and deriving the asymptotic behavior of $v(q | p ; C)$.
Building on this characterization, we propose a minimum-distance estimation approach for $\covar$ that accommodates multiple tail regimes. 
The methodology links CoVaR and $\Delta\covar$ to the underlying joint tail behavior, thereby providing a clear interpretation of these measures in systemic risk analysis. An empirical analysis across U.S. sectors demonstrates the practical value of the approach for assessing systemic risk contributions and exposures with important implications for macroprudential surveillance and risk management.
\end{abstract}

\noindent%
{\it Keywords:} conditional quantiles, copulas, tail order, tail dependence, extreme value theory
\vfill

\newpage
\spacingset{1.8} 

\section{Introduction}
\label{sec-intro}

From a macroprudential view \citep{imf2011gfsr,borio2012capital}, systemic risk arises from the gradual buildup of financial imbalances, such as credit expansion, rising leverage, and concentrated asset exposures.
While these imbalances do not immediately generate instability, they fundamentally alter how the system responds to severe shocks. In probabilistic terms, the buildup of systemic vulnerability manifests as changes in the extremal dependence structure of the system.

 The literature has developed a range of measures to quantify systemic risk by examining the system's response under stress \citep{noldeExtremeValueAnalysis2021}. Among them, conditional Value-at-Risk (CoVaR) \citep{adrian2016covar} is defined as the system's Value-at-Risk conditional on the distress of a given institution.
It has been extensively used in the literature to study systemic risk; see, e.g., \cite{fan2018single,ascorbebeitia2022effect}.

By its probabilistic nature, CoVaR is an
extreme conditional quantile.
The copula representation of CoVaR has been proven useful to study its theoretical properties; see, e.g., \cite{bernard2015conditional,li2026covar}. It makes explicit how the extremal dependence structure, captured by the tail behavior of the joint copula, determines CoVaR through an adjusted probability level.
This copula-adjusted level, denoted by $v(q \mid p; C)$, represents CoVaR on the uniform scale and lies at the core of our methodology.
Earlier works along this direction include \cite{jaworski2017conditional}
and 
\cite{nolde:2025}, 
which derived the limit of $v(q \mid p; C)$ as $p \downarrow 0$ under asymptotic dependence.

Previous work in the copula literature, e.g., \cite{hua:2011,li2023estimation}, has developed a tail expansion framework to describe the extremal dependence structure of multivariate random vectors. 
Specifically, a bivariate copula $C$ with a well behaved joint lower tail is said to admit a lower tail expansion if there exist a tail order function $a(w_1, w_2):(0, \infty)^2 \rightarrow (0, \infty)$, a tail order $\kappa \geq 1$, and a slowly varying function $\ell(u)$ such that, as $u \rightarrow 0^{+}$,
$$
a(w_1, w_2) = \lim_{u \downarrow 0} \frac{C(u w_1, u w_2)}{u^{\kappa}  \ell(u)}.
$$
The tail behavior is further refined based on the tail order: $C$ has strong tail dependence if $\kappa = 1$, intermediate tail dependence if $\kappa \in (1,2]$, tail orthant independence if $\kappa = 2$, and super tail orthant independence if $\kappa > 2$.
The recent work 
\cite{li2026covar} 
has established the value of tail expansions in analyzing the limiting behavior of $v(q \mid p;C)$ at the extreme quantile level $q=p$ for $\kappa \in [1,2]$.

In this work, we propose an extreme value inference framework for CoVaR centered on this copula-adjusted probability level. One main contribution is to characterize the possible tail regimes, corresponding to tail attraction, tail repulsion, and tail balance between firm-level distress and system-wide loss, through the limiting behavior of the copula conditional distribution. We then formally establish how these regimes are determined by tail expansions of the copula, thereby providing tractable conditions for regime identification and for deriving the behavior of $v(q \mid p;C)$.

Building on this characterization, we propose a minimum-distance estimation approach for $\covar(q \mid p)$ that accommodates multiple tail regimes and establishes its asymptotic consistency.
A similar approach was used in 
\cite{nolde:2025} 
to estimate $\covar(q \mid p)$ for the case $q = p$ under the regime that we refer to as strong tail attraction.

We also introduce a $\Delta \covar$ and derive its limiting behavior as $p \downarrow 0$.
Unlike the $\Delta \covar$ in earlier works such as \cite{adrian2016covar,girardi2013systemic}, our copula-based formulation isolates the effect of extremal dependence from that of marginal volatility. It aligns with the IMF’s macroprudential view \citep{imf2011gfsr} of systemic risk and supports monitoring tools that distinguish structural fragility from high-frequency market turbulence.

The rest of the paper is organized as follows.
 \cref{sec:2} introduces the basic definitions and some properties of $\covar$ and $\Delta \covar$. \cref{sec:3} and \cref{sec:4} present our main  results on the behavior of
  $\covar$ and $\Delta \covar$, respectively. \cref{sec:5} develops the estimation procedure and establishes its asymptotic properties.
\cref{sec:empirical} demonstrates the empirical value of the proposed methodology through a comprehensive analysis of systemic risk in the U.S. market, highlighting how extremal dependence within the system shapes $\covar$ and $\Delta \covar$ and reveals differences in the systemic roles of assets and institutions. \cref{sec:discussion} concludes with a brief discussion of the implications for macroprudential monitoring and risk management.
Proofs and technical details are provided in the supplementary material.

\section{Basic definitions and properties of CoVaR}
\label{sec:2}

Let $(X_i,X_S)$ denote, respectively, the return of institution $i$ and the
system-wide return. Their bivariate distribution is
$
F_{X_i,X_S}(x,s)=\Pr(X_i\le x,\;X_S\le s)$, for
$(x,s)\in\mathbb R^2,
$
with continuous marginals $F_{X_i}$ and $F_S$.
Define the generalized inverse $g^{\leftarrow}(u):=\inf \{x \in \mathbb{R}: g(x) \geq u\}$,
where $g^{\leftarrow}=g^{-1}=\{x:g(x)=u\}$ if $g$ is strictly monotone.
\begin{definition}
\label{def:covar}
Fix a small $p \in(0,1)$ and $\var_i(p)=F^{\gets}_{X_i}(p)$ be the individual $p$-level VaR.
Define the conditional distribution of $X_S$
given the stress event $\{X_i\le \var_i(p)\}$ by
\[
F_{X_S\,\mid\,X_i\le\var_i(p)}(s)
   =\Pr \bigl(X_S\le s \mid X_i\le\var_i(p)\bigr).
\]
The conditional Value-at-Risk of the system at level $q \in (0,1)$ is
\[
\covar_{S \mid i}(q\mid p)
:=F^{\gets}_{S\,\mid\,X_i\le\var_i(p)}(q).
\]
\end{definition}
The $
\covar_{S\mid i}(q\mid p)$ admits an equivalent copula representation.
Let $U_i=F_i\left(X_i\right)$ and $U_S=F_S\left(X_S\right)$. The associated copula $C_{iS}$, such that $(U_i, U_S) \sim C_{iS}$, is uniquely specified, for $(u,v)\in(0,1)^2$,
$
C_{iS}(u,v) =F_{X_i,X_S}\!\bigl(F^{\gets}_{X_i}(u),\,F^{\gets}_{S}(v)\bigr).
$
For each $p\in(0,1)$, define
\[
C_{S\mid i\le}(v\mid p):=\Pr(U_S\le v\mid U_i\le p)=\frac{C_{iS}(p,v)}{p},\qquad v\in[0,1].
\label{eq:conditional_cdf}
\]
Then 
\[
\covar_{S\mid i}(q\mid p)
=(F_S \circ C_{S\mid i\le}(\cdot\mid p))^{\gets}(q)
=\var_S\big(v_{S \mid i}(q \mid p;\, C_{iS})\big),
\label{eq:covar_general}
\]
and the copula-adjusted probability level is
\begin{align}
v_{S \mid i}(q \mid p;\, C_{iS}):=C_{S \mid i \leq}^{\leftarrow}(q \mid p)=\inf \left\{v \in[0,1]: C_{S \mid i \leq}(v \mid p) \geq q\right\}.
\label{eq:vqp_general}
\end{align}

This representation makes explicit how the dependence structure of $(X_i,X_S)$, encoded by their joint copula $C_{iS}$, will affect
$\covar$ via the adjusted probability level. In other words, the copula reflects how the system’s return quantiles shift under the stress event.
As emphasized in \cite{adrian2016covar,girardi2013systemic}, systemic impact is better reflected by the shift in CoVaR than by the CoVaR level itself. For this end, we introduce the risk measure $\Delta\covar$. Unlike the existing literature, we use the unconditional $\var$ as the benchmark for normal market conditions. As shown in \cref{sec:4}, this choice gives $\Delta \covar$ a clear interpretation in terms of the joint tail behavior of the risk variables and its effect on systemic risk.


\bdef[$\Delta$CoVaR]
\label{def:deltacovar}
\[
\Delta\covar_{S\mid i}(q\mid p)
:=\frac{\covar_{S\mid i}(q\mid p)-\var_S(q)}{|\var_S(q)|}
  =\frac{\var_S\!\bigl(v_{S\mid i}(q\mid p;C_{iS})\bigr)}
         {|\var_S(q)|}-1.
\]
\edef

Below are some basic properties of the copula-adjusted probability level, or $\covar$ in the uniform (0,1) scale. For the remainder, we omit the subscript when there is no ambiguity.

If $(X_i,X_S)$ has copula $C$, then $(-X_i,-X_S)$ has its reflected or survival copula $C^*(u,v)=u+v-1+C(1-u,1-v)$; $(-X_i,X_S)$ has 
$1$-reflected copula, 
$C^{ 1*}\left(u, v\right)=v-C\left(1-u, v\right) =u-C^*(u,1-v)$; 
$(X_i,-X_S)$ has its $2$-reflected copula, $C^{2* }\left(u, v\right)=u-C\left(u, 1-v\right)$.

\bprop
\label{prop:properties-of-v(q|p)}
\begin{enumerate}[label=(\alph*)]
\item Comonotonicity: $v(q \mid p; C^{+})=p q$.

\item Independence: $v(q \mid p; C^{\perp})=q$.

\item Countermonotonicity: $v(q \mid p;C^{-})=1-(1-q) p. $

\item Coherence:
if $C_1\preceq_c C_2$ (i.e.\ $C_1(u,v)\le C_2(u,v)$ for all $(u,v)$) then for all $p,q$,
      $ v(q\mid p;C_2)\;\le\;v(q\mid p;C_1)$.
     \item Reflection:
$ v(q\mid p;C)=
        1-v(1-q\mid p;C^{2*}), \quad
         v(q\mid p;C^*)=
        1-v(1-q\mid p;C^{1*})$.  
\item Range: 
$p q \leq v(q \mid p;C) \leq q
$ if  $C^{\perp} \prec_c C \prec_c C^{+}$ and
$ q \leq v(q \mid p;C) \leq 1-(1-q)p $ 
if $C^{-} \prec_c C \prec_c C^{\perp}$.
\end{enumerate}
\eprop

\section{Tail regimes of the copula-adjusted level in CoVaR}
\label{sec:3}

This section studies the limiting behavior of the copula-adjusted probability level $v(q \mid p ; C)$, which is central to the copula representation of CoVaR in \cref{eq:covar_general}. We show that, as $p \downarrow 0$, the limit of $v(q \mid p ; C)$ is determined by the limit laws of the conditional 
cumulative distribution function (cdf)
$C_{S \mid i \leq}(\cdot \mid p)$ which in turn induces a classification of tail regimes.
We then link these regimes to the joint tail behavior of the variables through the tail expansion of the copula, which yields a criterion to identify the tail regime and 
refined characterization of 
$v(q \mid p;C)$.
\cref{sec:3-1} and \cref{sec:3-2} treat, respectively, the fixed $q$ case and the extreme quantile case $q=p$.

\subsection{Tail regimes of $v(q\mid p;C)$ for fixed $q$}
\label{sec:3-1}

\bthm
\label{thm:limit_vqp}
Suppose there exists a function $A:(0,1)\to[0,1]$ such that, for $v\in(0,1)$,
\begin{align}
\lim_{p\downarrow0} C_{S\mid i\le}(v\mid p)=A(v).
\label{eq:Av}
\end{align}
Assume that the convergence is locally uniform on $(0,1)$, i.e., for every $0<a<b<1$,
\begin{align}
\sup_{v\in[a,b]}\bigl|C_{S\mid i\le}(v\mid p)-A(v)\bigr|\xrightarrow[p\downarrow0]{}0.
\label{eq:local_uniform_A}
\end{align}
Then the following hold:
\begin{enumerate}[label=\textup{(\roman*)}]
\item $A$ is non-decreasing
and continuous on $(0,1)$.
\item $A(v)$ extends to a proper distribution 
function supported on $[0,1]$ with possible jumps at $0$ and $1$ of sizes
$
p_0:=A\left(0^{+}\right)$, $ p_1:=1-A\left(1^{-}\right)$, $ 0 \leq p_0, p_1 \leq 1,  p_0+p_1 \leq 1 .
$
\end{enumerate}
With $v(q \mid p;C)$ defined in \cref{eq:vqp_general},
 for every $q \in(0,1)$,
$$
\lim _{p \downarrow 0} v(q \mid p ; C)= \begin{cases}0, & 0<q < p_0 \\ A^{\gets}(q) \in (0,1),
& p_0 < q < 1-p_1, \text{ provided $A^{\leftarrow}$ is continuous at $q$}\\ 1, & 1-p_1 < q<1\end{cases}.
$$
\ethm

\brmk
The boundary cdf $\lim_{u \to 0}\Pr\left(U_S \leq v \mid U_i = u\right)$
has been studied in \cite{hua2014strength}; it assumes that the copula
has a continuous first-order derivative with respect to $u$.
The boundary of $C_{S \mid i \leq}(v \mid u)$
 has not, to our knowledge, been studied before.
It does not require that the copula have continuous derivatives.
When the copula is continuously differentiable in $u$, the two boundary cdfs coincide in the limit as $u \rightarrow 0$, following from l'Hopital's rule.
\ermk

\cref{thm:limit_vqp} characterizes the possible tail regimes of $v(q \mid p ; C)$ based on the structure of the limiting conditional distribution $A$.
\begin{enumerate}[label=\textup{(\roman*)}]
\item Tail attraction\\
If $p_0=1$ and $p_1=0$, then the extended distribution of $A(v)$ is a pure atom at $0$. Consequently,
\[
v(q \mid p; C) \to 0, \qquad q\in(0,1).
\]
We refer to this regime as tail attraction, as the extremely small values of $U_i$
pull $U_S$ towards the lower endpoint $0$.

\item Tail repulsion\\
If $p_0=0$ and $p_1=1$, then the extended distribution of $A(v)$ is a pure atom at $1$. Consequently,
$$
v(q \mid p; C) \to 1, \qquad q\in(0,1).
$$
We refer to this regime as tail repulsion, as conditioning on extremely small values of $U_i$  pushes $U_S$ away from $0$ and toward the opposite endpoint $1$.

\item Tail balance\\
If $p_0=p_1=0$, then the extended  distribution of $A(v)$ has no boundary atoms, so all mass lies entirely in the interior. In this case,
\[
v(q \mid p;C) \to A^{\leftarrow}(q)\in(0,1).
\]
We call this regime tail balance, as extreme values of $U_i$ do not force $U_S$ to the boundary but balance it in the interior.

\item Mixed tail regimes\\
We refer to the case as mixed tail regimes when the extended limiting distribution has at least one boundary atom and is not a single point mass. It places mass $p_0$ at $0$, mass $p_1$ at $1$ and the remaining $1-p_0-p_1$
on the interior.
The behavior of $v(q \mid p ; C)$ exhibits mixed regimes depending on how the total mass is distributed between the two boundary atoms and the interior component.
\end{enumerate}

A central contribution of this paper is to link the behavior of the limiting conditional distribution $A(v)$ with the joint tail expansion of the copula 
so that we can characterize the asymptotic regimes of the quantiles based on tail properties of the copula, i.e., the tail order and the tail dependence function.
In
\cref{prop:kappaone}, 
we showed that when the copula has strong lower tail dependence with $\kappa=1$, the conditional limit $A(v)$ will have a strictly positive mass at $0$ with size equal to the limit of its tail dependence function.



\bprop
\label{prop:kappaone}
Assume the copula $C$ admits a lower tail expansion with a tail order $\kappa = 1 $, as $u \downarrow 0$,
\begin{align}
\frac{C(uw_1,u w_2)}{u}= b(w_1,w_2) + R(u; w_1, w_2),
\label{eq:tail_expansion_b}
\end{align}
in which  $R\left(u ; w_1, w_2\right) \rightarrow 0$ uniformly whenever $u w_1 \rightarrow 0$ and $u w_2 \rightarrow 0$.
The tail dependence function $b(w_1,w_2)$ is monotone and bounded with 
$
b_{\infty}:=\lim _{r \rightarrow \infty} b(1, r)=\sup _{r>0} b(1, r) \in (0,1] .
$
Then the function $A$ defined in \cref{eq:Av} satisfies
$$
A(0^+)=b_{\infty} \in (0,1].
$$
\eprop

The results in the following lemma aim to relate the behavior of $v(q \mid p; C)$ to the tail expansion of the copula and its 2-reflected copula. The proof follows from the definition of $C^{2*}$.

\begin{lemma}
\label{lem:2ref_equivalence}
Let $C$ be a bivariate copula. Define its $2$-reflection
\[
C^{2*}(u,v):=u-C(u,1-v), \qquad (u,v)\in[0,1]^2,
\]
which is the copula associated with the random vector $(X_i,-X_S)$.
The conditional distribution function
\[
C^{2*}_{S\mid i\le}(v\mid p):=\Pr(1-U_S\le v \mid U_i\le p)=1-C_{S\mid i\le}(1-v\mid p).
\]
Assume that for every $v\in(0,1)$, 
$
A(v)=\lim_{p\downarrow0} C_{S\mid i\le}(v\mid p)
$
exists. Then 
$
A^{2*}(v):=\lim_{p\downarrow0} C^{2*}_{S\mid i\le}(v\mid p)
$
also exist and satisfy
\[
A^{2*}(v)=1-A(1-v), \qquad v\in(0,1).
\]
\end{lemma}

\bthm
\label{thm:limit_vqp_tail}
Assume the copula $C$ admits a lower tail expansion with tail order $\kappa\ge 1$,
\begin{align}
\frac{C(uw_1,uw_2)}{u^\kappa}
= a(w_1,w_2)\,\ell(u)+R(u;w_1,w_2), \qquad u\downarrow0,
\label{eq:tailex_remainder}
\end{align}
where $\ell$ is slowly varying at $0$,  $R\left(u ; w_1, w_2\right) \rightarrow 0$ uniformly whenever $u w_1 \rightarrow 0$ and $u w_2 \rightarrow 0$.
When $\kappa=1$, the expansion reduces to \cref{eq:tail_expansion_b} and the tail order function $a$ 
coincides with the tail dependence function $b$.

Assume in addition, the following boundary conditions on the tail order function,
\begin{enumerate}[label=\textup{(B\arabic*)}]
    \item \label{it:B1}
    As $r\downarrow0$,
    $ a(1,r)=a_1 r+o(r)$
    for some constant $a_1>0$.

    \item \label{it:B2}
If $\kappa=1$, let $H(r) = a(1,r)$, which is continuous and strictly increasing on $(0,\infty)$, with
$
\lim_{r\downarrow0}H(r)=0$, $
\lim_{r\to\infty}H(r)=b_\infty\in(0,1].
$
If $\kappa>1$, then $a(1,r)=O(r^\rho)$ for some $0<\rho\le \kappa-1$ as $r \to \infty$.
\end{enumerate}

Then the following hold.
\begin{enumerate}[label=\textup{(\roman*)}]
    \item   
    The limiting conditional cdf $A$ has a strictly positive mass at $0$,
    i.e.
    $
    A(0^+)>0,
    $
    if and only if $C$ has strong lower tail dependence, i.e., $\kappa=\kappa(C)=1$. Moreover, when $\kappa=1$, $A(0^+)=b_{\infty}$ and    
    for every $q\in(0,b_\infty)$,
    \[
    v(q\mid p;C)\sim H^{-1}(q)\,p \downarrow 0 .
    \]

    \item    
  The limiting conditional cdf $A$ has a strictly positive mass at $1$,
    i.e.
    $
    1-A(1^-)>0,
    $
    if and only if the $2$-reflected copula $C^{2*}$ has strong lower tail dependence, i.e., $\kappa^{2*}=\kappa(C^{2*})=1$. In that case, $
    1-A(1^-)=b_\infty^{2*},$
    and for every $q\in(1-b_\infty^{2*},1)$,
    \[
    v(q\mid p;C)\sim 1-(H^{2*})^{-1}(1-q)\,p \uparrow 1.
    \]
\end{enumerate}
\ethm

\brmk
The conclusion of the theorem relies on two technical conditions, the uniform expansion of the copula and growth rate of the tail order function. If the conditions fail, the tail order $\kappa$ alone no longer determines the boundary behavior of the limiting conditional cdf. It is possible to have $\kappa>1$ while 
$A(0^+)>0$ and the conditional quantiles converge at non-polynomial rates that are not captured by the expansion used in the theorem.
The Gumbel copula and the Gaussian copula are two examples 
with tail order $\kappa>1$ and $A$ in \cref{eq:Av} degenerate at 0.
For these two families, the tail expansion in \cref{eq:tailex_remainder} is not uniform; the remainder in the expansion inflates away from the diagonal.
\ermk

\begin{table}[ht]
\renewcommand{\arraystretch}{1.5}
\centering
\resizebox{\textwidth}{!}{%
\begin{tabular}{lllll}
\hline
\textbf{Copula} & $\kappa$ & $A(v)$  & $v(q \mid p)$ & $v(p)$ \\
\hline
Clayton($\theta>0$) & 1 & 1 & $p(q^{-\theta}-1)^{-1 / \theta}$ & $p^2$ \\
Gumbel$^*$($\delta>1$)&
1
&1 & $H^{-1}(q;\delta) p$
& $p^2$\\
IPS$^{*}$ ($\theta>0$) & $1+\dfrac{1}{\theta}$ &
$1-e^{-F_{\Gamma}^{-1}(v ; \theta)}$ &
$F_{\Gamma}(-\ln(1-q);\theta)$ &
$(\Gamma(\theta+1))^{-1/\theta}p^{\,2-1/\theta} \,(\theta>1)$\\
& & & & $\{\Gamma(\theta+1)\}^{-1} p^\theta \, (0<\theta<1)$\\
Frank ($\theta>0$)& 2& $(1-e^{-\theta v})(1-e^{-\theta})^{-1} $ &
$-\frac{1}{\theta} \log (1-q(1-e^{-\theta}))$ &$\theta^{-1}(1-e^{-\theta}) p$\\
Gumbel$^{2*} (\delta>1)$&
$1+\delta$&
0&
$1-\exp \left\{-(\delta q)^{1 / \delta}(-\log (p))^{1-\frac{1}{\delta}}\right\}$
&$(\delta p)^{1 / \delta}(-\log (p))^{1-\frac{1}{\delta}}$\\
Clayton$^{2*} (\theta>0)$& $\theta+2$ & 0& $1-((1-q)^{-\theta}-1)^{-1 / \theta} p$
& $1-\theta^{-1 / \theta} p^{1-1/\theta}\,(\theta>1)$\\
& & & & $p^{1-\theta} \,(0<\theta<1)$\\
\hline
\end{tabular}
}
\caption{ Tail order $\kappa$, limiting conditional cdf $A(v)$, copula-adjusted level $v(q \mid p; C)$ and $v(p;C)$
for common parametric copula families. A superscript $*$ denotes the reflected copula, $1*$ the 1‑reflected, and $2*$ the 2‑reflected copula. The IPS copula is an Archimedean copula constructed from the integrated positive stable Laplace transform; see Section 4.11 of \cite{joe2014dependence} for details.}
\label{tab:copula_expansions}
\end{table}

\cref{thm:limit_vqp_tail} clarifies the link between the behavior of $A$,
$v( q\mid p;C)$ and the tail expansion of the copula. 
When the copula $C$ has tail order $1$, the tail mass is asymptotically concentrated in the lower left corner.  The limiting conditional cdf $A$ has positive mass at $0$, and the conditional quantiles converge toward $0$ at a linear rate in $p$.
Conversely, if the 2-reflection of $C$ has tail order $1$, then the tail mass is asymptotically concentrated in the upper left corner. In that case, $A$ assigns positive mass to $1$, and the conditional quantiles converge toward $1$ at a linear rate in $p$.

Most common copula families fall into a single regime for $v(q
\mid p; C)$.
\cref{tab:copula_expansions} summarizes the tail order $\kappa$, the limiting conditional cdf $A(v)$, and the behavior of $v(q
\mid p; C)$ for common copula families.
The first two rows, corresponding to the Clayton and reflected Gumbel copula families, both exhibit strong lower tail dependence: $A$ is degenerate at $0$, there is tail attraction for all $q \in (0,1)$, with $v(q
\mid p; C) \to 0$ at a linear rate.
The middle two rows, corresponding to the reflected IPS and Frank copula
families. 
In both cases $\kappa > 1$, $A(v)$ is strictly increasing, and $v(q
\mid p; C) \to A^{-1}(q)$ for all $q \in (0,1)$, reaching a tail balance.
The last two rows correspond to $2$-reflected copula families, 
where the reflection operation converts families with only positive
dependence to families with only negative dependence.
In these cases, $A(v)$ is degenerate at $1$, and there is tail repulsion: $v(q
\mid p; C) \to 1$ for all $q \in (0,1)$.

The reflected IPS copula family has cdf
  $$C^{2*}(u,v;\theta) 
  = u+v-F_\Gamma\bigl((x^\theta+y^\theta)^{1/\theta};\theta\bigr),
  \quad x=F_\Gamma^{-1}(u;\theta), \ y=F_\Gamma^{-1}(v;\theta),\ \theta>0,$$
where $F_\Gamma(\cdot;\theta)$ and $F_\Gamma^{-1}(\cdot;\theta)$ are the cdf
and quantile functions for the Gamma$(\theta,1)$ distribution.
It has negative dependence for $0<\theta<1$, independence for $\theta=1$
and positive dependence for $\theta>1$.
The tail order is $\kappa(\theta)=1+\theta^{-1}>1$. 
This family is useful for $\kappa>1$ in (B2) of \cref{thm:limit_vqp_tail}.
It can be shown directly that its tail order function in \cref{eq:tailex_remainder} is
$a(w_1,w_2) = (w_1+w_2)^{\kappa} - w_1^{\kappa} -w_2^{\kappa}$
with $\ell(u) = [\Gamma(\theta+1)]^{1/\theta}/\kappa$.
Then $a(1,r)\sim \kappa r^{\kappa-1}$ as $r\to\infty$,
so that the condition in (B2) is satisfied with $\rho=\kappa-1$.

Mixed tail regimes arise when the  tail mass is split between the lower and upper corners. This is less common among standard parametric copula families.
The Student‑t copula has tail order 1 in all four corners, which results in concentration of tail mass in both the lower and upper corners, and thus exhibits a mixed tail regime.

\bexa
\label{exa:student_t}
Let $C_{\rho, \nu}$ be the bivariate Student t$_\nu$ copula with correlation $\rho \in(-1,1)$ and degrees of freedom $\nu>0$. Its lower tail dependence function is
$$
b\left(w_1, w_2 ; \rho, \nu\right)=w_1 T_{\nu+1}\big\{K\big[\rho-(\frac{w_2}{w_1})^{-1/\nu}\big]\big\}
+w_2 T_{\nu+1}\big\{K\big[\rho-(\frac{w_1}{w_2})^{-1/\nu}\big]\big\},
$$
where $T_{\nu+1}$ is the univariate $t$ CDF with $\nu+1$ degrees of freedom and
$
K=\sqrt{(\nu+1) /\left(1-\rho^2\right)}.
$
Let $H(r)=b(1,r;\rho,\nu)$ and $H^{2*}(r)=b(1,r;-\rho,\nu)$.

For $v \in (0,1)$, 
$
A(v)=\sup_{r>0}H(r)= T_{\nu+1}\left(\rho K\right)= q^*,
$
and has a jump of size $p_1=1-q^*$ at $1$. It thus exhibits a mixed tail regime depending on $q$.
For $q< q^*$, it has tail attraction
$v(q \mid p; C) \sim H^{-1}(q) p \to 0.
$
For $q > q^*$, it has tail repulsion,
$
v(q \mid p; C) \sim 1-(H^{ 2*})^ {-1}(1-q) p
\to 1.$
 For $q=q^*$, 
$v(q^* \mid p;C) \to \frac{1}{2}.$
\eexa

\subsection{Tail regimes of $v(q\mid p;C)$ for extreme-level $q$}
\label{sec:3-2}

In systemic risk measurement, we are particularly interested in extreme quantile levels, typically with the same tail probability used for both the institution and the system, i.e., $q=p$. Under this setting, \cref{def:covar} simplifies to:
$$
\covar_{S \mid i}(p)= \var_{S}(v(p;C_{iS})),
$$
and the adjusted level 
$
v(p;C_{iS})=C_{S \mid i \leq}^{\gets}(p \mid p).
$
A key difference is that, under $q=p$, the right-hand-side target probability becomes smaller order than $p$. Driving $v(p)$ to the lower endpoint, as a result, does not require the same strength of tail concentration as in \cref{sec:3-1}.  \cref{thm:limit_vp} derives the limit of $v(p)$ based on the limiting conditional cdf, showing that $v(p) \rightarrow 0$ holds broadly except in the degenerate case where mass concentrates in the upper left corner, in which case higher-order terms determine the limit.

\bthm
\label{thm:limit_vp}
With $A(v)$ defined in \cref{eq:Av},
\begin{enumerate}[label=\textup{(\roman*)}]
    \item If $A(0^+)=p_0>0$, as $p \downarrow 0$, 
    $$
    v(p;C) \to 0.
    $$
    \item If $A(0^+)=p_0=0$ and $A(1^-)=1-p_1 \in (0,1]$, and $v_{-}=\inf \{v \in[0,1]: A(v)>0\}=0$, then
    as $p \downarrow 0$, 
    $$
    v(p;C) \to 0.
    $$ 
    \item   If $p_0=0$, $p_1=1$, then
    $A(v)=0$ for all $v \in(0,1)$, it would depend on the higher-order terms. Assume there exists $\delta>0$,
  a non-decreasing function $B:(0,1) \rightarrow (0, \infty)$, and $\ell$ is slowly varying with $\ell_0:=\lim _{p \downarrow 0} \ell(p) \in(0, \infty)$ such that 
    uniformly for $v \in (0,1)$
\begin{equation}
    C(p, v)=B(v)p^{1+\delta}\ell(p)+o(p^{1+\delta}). \label{eq:th3.8(iii)}
\end{equation}
     \begin{enumerate}[label=\textup{(\alph*)}]
      \item $0<\delta<1$: $v(p;C)\to 0$.
       \item $\delta=1$: $v(p;C)\to B^{\gets}(\ell_0^{-1})$ if $B^{\leftarrow}$ is continuous at $\ell_0^{-1}$.

      \item $\delta>1$: $v(p;C) \to 1$.  
      \end{enumerate}
\end{enumerate}
\ethm

The next theorem characterizes the behavior of $v(p ; C)$ using the lower tail expansion of the copula. It provides a criterion in terms of the tail order and yields explicit convergence rates for $v(p ; C)$. The case of $\kappa \in [1,2]$ has been studied in 
\cite{li2026covar}.

\bthm
\label{thm:limit_vp_tail}
Assume the copula $C$ admits the lower-tail expansion as described in \cref{thm:limit_vqp_tail}. Then
$$
v(p ; C) \to 0 \Longleftrightarrow 1 \leq \kappa<2+\rho.
$$
Moreover as $p \downarrow 0$,
\begin{align}
v(p;C)= \begin{cases}
O(p^{3-\kappa}), 
& 1 \leq \kappa \leq 2, \\ 
O(p^{1-\frac{\kappa-2}{\rho}}), 
& 2<\kappa<2+\rho .\end{cases}
\label{eq:limit_vp}
\end{align}
\ethm

The last column of \cref{tab:copula_expansions}
shows $v(p; C)$ for common copula families.
Clayton and reflected Gumbel copulas have $\kappa = 1$, so $v(p) = O(p^2)$.
The reflected IPS copula has  tail order
$\kappa = 1 + 1/\theta $ 
and $\rho=\kappa-1$ for (B2) of \cref{thm:limit_vqp_tail}; then  
$v(p) = O(p^{2-1/\theta})$ when $\theta>1$ ($1<\kappa\le2$)
and $v(p) = O(p^{1/(\kappa-1)})= O(p^\theta)$ 
when $0<\theta<1$ ($2<\kappa<2+(\kappa-1)$).
The Frank copula has tail orthant independence with $\kappa = 2$
for all $-\infty<\theta<\infty$, and hence $v(p) = O(p)$.
The 2-reflected Gumbel and Clayton copulas exhibit negative quadrant dependence with $\kappa > 2$.
For the former, it has $2<\kappa < 2 + \rho$ for all $\delta>1$, and
$
v(p) = O(p^{1/\delta}(-\log p)^{1 - 1/\delta}) \downarrow 0.
$
For the latter,  $2<\kappa < 2 + \rho$ when $0 < \theta < 1$, so
$
v(p) = O(p^{1 - \theta}) 
$
when $0<\theta<1$.
However, when $\theta > 1$, we instead have $v(p) \sim 1-\theta^{-1 / \theta} p^{(\theta-1) / \theta}\uparrow 1$.

\section{Tail limits of $\Delta$CoVaR and systemic risk implications}
\label{sec:4}

In quantifying systemic risk impact, the primary concern is its effect on extreme system losses. Therefore, $\Delta \operatorname{CoVaR}_{S \mid i}(q \mid p)$ is most informative when evaluated at small $q$. When $q=p$, \cref{def:deltacovar} reduces to
\[
\Delta\!\operatorname{CoVaR}_{S\mid i}(p)
     \;=\;
     \frac{\var_S(v(p;C))-\var_S(p)}          {|\var_{S}(p)|} ,
\]
which measures the percentage change in the system’s VaR under distress conditions relative to normal conditions.
Its behavior will depend on the 
the limiting behavior of $v(p)$, determined by the tail dependence structure of the copula, and $\var$, determined by the marginal distribution of $X_S$, as detailed in the following theorem.

\bthm
\label{thm:deltaCoVaR}
Let $(X_i,X_S)$ be the returns of institution $i$ and the system, with joint cdf  $
F_{X_i,S}(x,s)=\Pr (X_i\le x,\;X_S\le s)
=C_{iS}(F_{X_i}(x),F_{X_S}(s)).
$
Assume that 
\begin{enumerate}[label=\textnormal{(A\arabic*)}]
\item $F_S$ lies in the minimum‐domain of attraction of a univariate 
generalized extreme value (GEV) distribution with lower–tail index $\xi= \xi_S \ge 0$. Hence, as $p \downarrow 0$
$$
\var_S(p) \sim \begin{cases}-p^{-\xi} L^*(1 / p), & \xi>0, \\ -H^{-1}(-\log p), & \xi=0.\end{cases}
$$
where $L^*$ is slowly varying at $\infty$ and $H^{-1}$ has extended regular variation at $\infty$ with index $\gamma \in (0, \infty]$.

\item The copula $C_{iS}$ admits a lower expansion of \cref{eq:tailex_remainder}, with tail order $\kappa=\kappa_{iS} \in [1,2+\rho)$ so that $v(p;C)$ behaves as described in \cref{thm:limit_vp_tail}.

\end{enumerate}

Then, as $p\downarrow0$, the following hold:

\begin{enumerate}[label=\textup{(\roman*)}]
\item Heavy marginal tail $(\xi>0)$
$$
\Delta \covar_{S \mid i}(p) \sim \begin{cases}-p^{-(2-\kappa) \xi} \to -\infty, & 1 \leq \kappa<2, \\ 1-a_0^{-\xi} \in (-\infty,1), & \kappa=2,\quad\text{with } a(1,a_0)=1. \\ 1-p^{\frac{(\kappa-2)\xi}{\rho}} \to 1, & 2<\kappa<2+\rho .\end{cases}
$$

\item Light marginal tail $(\xi=0)$
$$
\Delta \covar_{S \mid i}(p)\to \begin{cases}
1-(3-\kappa)^{\gamma} < 0, & 1 \leq \kappa<2, \\ 0,
& \kappa=2,\\ 1-(1-\frac{\kappa-2}{\rho})^{\gamma} \in (0,1], & 2<\kappa<2+\rho.\end{cases}
$$
\end{enumerate}
\ethm

This result provides a clear interpretation of $\Delta \operatorname{CoVaR}$
based on the joint tail behavior of the distribution, through the 
 copula tail and the marginal 
tail.

When the copula exhibits strong to intermediate tail dependence (i.e., $\kappa \in [1,2)$), 
$
\Delta \operatorname{CoVaR}_{S \mid i}(p)
$ is eventually negative.
This reflects a risk amplification effect: under the distress of institution $i$, the system is pushed deeper into the tail.
The amplification factor is determined by the marginal tail distribution.
If $F_S$ lies in the Fr\'echet  domain $(\xi>0)$, the amplification scales as $-p^{-(2-\kappa) \xi}$ and diverges as $p \downarrow 0$. If $F_S$ lies in the Gumbel domain $(\xi=0)$, the amplification converges to the finite limit $1-(3-\kappa)^\gamma$, with larger $\gamma$ (faster growth of $H^{-1}$ ) producing a stronger effect.

When $2<\kappa<2+\rho$,
which can occur with negative dependence,
$\Delta \operatorname{CoVaR}_{S \mid i}(p)$ is asymptotically positive.
This reflects a risk attenuation effect:
under the distress of institution $i$, the system’s tail losses become less extreme. The attenuation factor is between $0$ and $1$, and tends to $1$ when $\xi>0$.
Note,  when $v(p) \to 1$,
the factor can exceed $1$
or even diverges to $\infty$.

In the middle case with tail-orthant independence $(\kappa=2)$, if $\xi=0$ then $
\Delta \operatorname{CoVaR}_{S \mid i}(p) \rightarrow 0,
$
indicating that the system's tail risk is asymptotically unchanged.
If $\xi>0$, the sign depends on the value of $a_0=\lim_{p \to 0}r(p;C)$.
$\Delta \operatorname{CoVaR}_{S \mid i}(p)$ is asymptotically negative when $a_0<1$ and 
 positive when $a_0>1$.

\section{Estimation of CoVaR and its asymptotic properties}
\label{sec:5}

Estimation of CoVaR, based on the copula representation in \cref{eq:covar_general}, involves estimating the marginal Value-at-Risk (VaR) and the copula-adjusted level 
$v(q \mid p;C)$. In this work, we propose a minimum-distance estimation approach for
$
v(q \mid p; C) 
$,
which represents CoVaR on the uniform scale, and can then be combined with the marginal distribution to estimate CoVaR on the original scale.

\subsection{Empirical estimators of tail dependence}
\label{sec:empirical_tail_funcational}

Let $\{(X_i,Y_i), i \in \{1,2,\ldots,\}\}$ be an infinite sequence of independent random vectors with common bivariate distribution function $F$ and univariate continuous margins $F_X, F_Y$.
Denote by $X_{i:n}$ be the $i$-th smallest among the first 
$n$ random variables $\{X_1, \ldots, X_n\}$. Define the transformed variables 
$U_i =F_{X}(X_i)$, and $V_i= F_{Y}(Y_i)$ and their empirical distribution function is 
$$
C_n(u, v)=\frac{1}{n} \sum_{i=1}^n \mathbf{1}\left\{U_{i} \leq u, V_i \leq v\right\}, \quad u, v \in (0,1) .
$$
In practice, $U_i$ and $V_i$ are unobserved, we replace them with pseudo-observations from ranks,
$$
U_{i,n}=
n^{-1}\sum_{j=1}^n \mathbf{1}\left(X_j \leq X_{i }\right),\quad
V_{i,n}=
n^{-1}\sum_{j=1}^n \mathbf{1}\left(Y_j \leq Y_{i }\right).
$$
Using these pseudo-observations, the empirical copula function is 
$$
\widehat{C}_n(u, v)=\frac{1}{n} \sum_{i=1}^n \mathbf{1}\left\{U_{i,n} \leq u, V_{i,n} \leq v\right\}, \quad u, v \in (0,1) .
$$

\bprop
Let $C_{XY}$ be the bivariate copula
of $(U_i, V_i)=(F_{X}(X_i), F_{Y}(Y_i))$. The conditional cdf is 
$ C_{Y \mid X\leq}(v\mid p)=C_{XY}(p,v)/p$, for $v \in (0,1)$.
Let $p_n=k_n/n$, and define the sequence of empirical estimators,
$$
\widehat{A}_{k_n,n}(v):=\frac{\widehat{C}_n\left(p_n, v\right)}{p_n}=\frac{1}{k_n}\sum_{i=1}^n \mathbf{1}\big(U_{i,n} \leq \frac{k_n}{n}, V_{i,n} \leq v\big), \quad  v \in (0,1).
$$
Suppose there exists a function $A:(0,1) \rightarrow(0,1)$ such that as $p \downarrow 0$,
$$
\sup _{v \in(0,1)}\left|C_{Y \mid X\leq}(v\mid p)-A(v)\right| {\longrightarrow} 0.
$$ 
Then as $n \to \infty$, with 
$k_n \to \infty$, $ {k_n}/{n} \to 0$ ,
${k_n}/{\sqrt{n}} \longrightarrow \infty$,
$$
\sup _{v \in(0,1)}\left|\widehat{A}_{k_n, n}(v)-A(v)\right| \xrightarrow{P} 0 .
$$
\label{prop:uniform_conv_Akv}
\eprop

\bprop
\label{prop:uniform_consistency}
Suppose the copula $C_{XY}$ admits a lower tail expansion in \cref{eq:tailex_remainder} with tail order $\kappa=1$ and tail dependence function $b(w_1,w_2;C)$.

With $p_n=k_n/n$, for fixed $\left(w_1, w_2\right) \in \mathcal{S}_1:=\left\{\left(w_1, w_2\right): w_1>0, w_2>0, w_1+w_2=2\right\}$, let
\begin{align}
\widehat{b}_{k_n, n}\left(w_1, w_2\right):=\frac{\widehat{C}_n\left(p_n w_1, p_n w_2\right)}{p_n}
=\frac{n}{k_n}\sum_{i=1}^n \mathbf{1}\Bigl(U_{i, n} \leq \frac{k_n}{n} w_1, V_{i, n} \leq \frac{k_n}{n} w_2\Bigr).
\end{align}
 Then as $n \to \infty$, $k_n/n \to 0$, $k_n \to \infty$,
\[
   \sup_{(w_1,w_2)\in\mathcal S_1}
      \bigl|\,
         \widehat{b}_{k_n,n}(w_1,w_2)
         \;-\;
      b(w_1,w_2;C)
      \bigr|
   \;\xrightarrow{\;P\;}\; 0.
\]
\label{prop:uniform_consistency_bhat}
\eprop

\subsection{Minimum distance estimation of $\covar$}
\label{sec:5-2}

The proposed estimation procedure is based on the classical principle of minimum distance estimation \citep{parr1982minimum}, where one fits a parametric model by aligning it as closely as possible to an empirical counterpart.
A similiar idea has been used in
\cite{einmahl:2012} and 
\cite{nolde:2025} 
for extreme value estimation.

Following this general principle, we define a criterion function $Q_n(\theta)$ as the integrated distance between an empirical tail functional and its model-based counterpart, and estimate the parameter $\theta$ by minimizing $Q_n(\theta)$.
The choice of the tail functional is guided by the theoretical results established in \cref{sec:3}, which
provides a basis to determine the regimes of $v(q \mid p;C)$ based on the tail order in the lower and upper left corner, $\kappa$ and $\kappa^{2*}$.  

With the assumptions in \cref{thm:limit_vqp_tail},
we consider the following cases.
\begin{enumerate}[label=\textup{(\roman*)}]
\item Tail attraction\\
If $\kappa=1$ and $\kappa^{2*}>1$, the tail mass is concentrated in the lower left corner. We assume a parametric model for the tail dependence function in the lower left corner $b(w_1,w_2; \theta)$ with $H(r)=b(1,r;\theta) \to 1$ as $r \to \infty$, and  construct the criterion function
\begin{align}
Q_{k_n, n}(\theta)=\int_{ \mathcal{S}_1}\left|\widehat{b}_{k_n,n}\left(w_1, w_2\right)-b\left(w_1, w_2 ; \theta\right)\right| \mathrm{d} w_1 \mathrm{~d} w_2 ,
\label{eq:Q_alpha_n}
\end{align}
where $\widehat{b}_{k_n,n}(w_1, w_2)$ is the empirical tail dependence function defined in \cref{prop:uniform_consistency_bhat}.
By \cref{thm:limit_vqp_tail},  the conditional quantiles can be estimated as, for $q \in (0,1)$,
\begin{align}
\hat{v}_n\left(q \mid p_n\right)=H^{-1}(q ; \hat{\theta}_n) \cdot p_n ,
\label{eq:v_q_hat}
\end{align}
which converges to $0$ as $p_n \to 0$.
\item Tail repulsion\\
If $\kappa>1$ and $\kappa^{2*}=1$, the tail mass is concentrated in the upper left corner, and we will assume a parametric model for $b(w_1, w_2 ; C^{2 *})$. The criterion function will be constructed based on $\widehat{b}_{k_n,n}\left(w_1, w_2\right)$ of the $2$-reflected data.
By \cref{thm:limit_vqp_tail}, the conditional quantiles can be estimated by 
\begin{align}
\hat{v}_n\left(q \mid p_n\right)=1-(H^{2*})^{-1}(1-q ; \hat{\theta}_n) \cdot p_n ,
\end{align}
which converges to $1$ as $p_n \to 0$.
\item Tail balance\\
If $\kappa>1$ and $\kappa^{2*}>1$, then by \cref{thm:limit_vqp_tail}, $A(v)$ does not have any boundary atoms. We therefore assume a proper parametric distribution $A(v; \theta)$ and define the criterion function as
\begin{align}
Q_{k_n, n}(\theta)=\int_0^1\left|\widehat{A}_{k_n, n}(v)-A(v ; \theta)\right| \mathrm{d} v,
\label{eq:Q_kn}
\end{align}
where $\widehat{A}_{k_n, n}$ is the empirical estimator for the conditional cdf defined in \cref{prop:uniform_conv_Akv}. The conditional quantile at level $q$ is then estimated by inverting the fitted model:
\begin{align}
\hat{v}_n\left(q \mid p_n\right)=A^{\gets}(q ; \hat{\theta}_n) .
\label{eq:vhat_A}
\end{align}
\end{enumerate}
The three cases above fall into a single tail regime, which covers the behavior of most parametric copula families. As discussed in \cref{sec:3}, however, mixed tail regimes may occur; for example, a tail order of 1 in both corners indicates a mixed regime. 
In this case, one could model using the first-order expansion of Student-$t$ copula, as in \cref{exa:student_t}, and construct a criterion function that targets both corners,
\begin{align}
Q_{k_n, n}(\rho, \nu)=\int_{\mathcal{S}_1}\left|\widehat{b}_{k_n,n}(w_1, w_2)\!-\!b(w_1, w_2 ; \rho, \nu)\right| 
+\left|\widehat{b}_{k_n,n}^{2*}(w_1, w_2)\!-\!b(w_1, w_2 ; -\rho, \nu)\right|
\mathrm{d} w_1 \!\mathrm{~d} w_2 .
\end{align}

The next two theorems establish the consistency of
$\hat{v}_n\left(q \mid p_n\right)$, the estimator of CoVaR on the uniform scale.
To recover $\covar$ on the original scale, we can plug it into a marginal quantile estimator, such as
$$
\widehat{\covar(q \mid p_n)}=\widehat{F}_{S,n}^{-1}\big(\hat{v}_n\left(q \mid p_n\right)\big),
$$
whose consistency follows from the continuous-mapping theorem, once the marginal quantile estimator is uniformly consistent on a neighbourhood of $v(q \mid p ; C)$.


\bthm
\label{thm:consistency_vqhatA}
Assume the model $A(\cdot;\theta)$ for the boundary cdf is correctly
specified. 
\begin{enumerate}[label=\textnormal{(A\arabic*)}]
  \item 
  \label{ass:compact}
        The parameter set $\Theta \subset \mathbb{R}^{p}$ is non-empty, closed, and bounded
        (hence compact).

  \item 
  \label{ass:uniform-tail} 
        There exists a true parameter value $\theta_0\in\Theta$ such that
        \[
           \lim_{p\downarrow 0}
           \frac{C(p,v)}{p}
           \;=\;
           A(v;\theta_0),
           \qquad\text{uniformly in }v\in(0,1).
        \]

  \item \label{ass:continuity}  
        For every $\theta\in\Theta$, the function $v\mapsto A(v;\theta)$ is a
    proper distribution function on $[0,1]$, continuous and
    strictly increasing on $(0,1)$.
    Moreover, the mapping $(v,\theta)\mapsto A(v;\theta)$ is jointly
    continuous on $[0,1]\times\Theta$.

  \item \label{ass:L1-id}
        For every $\varepsilon>0$,
        \[
           \inf_{\lVert\theta-\theta_0\rVert\ge\varepsilon}
           \int_{0}^{1}
           \bigl|A(v;\theta)-A(v;\theta_0)\bigr|\,dv
           \;>\;0.
        \]
\end{enumerate}

Let $\hat\theta_n$ be the minimizer of the criterion function in \cref{eq:Q_kn},
and the plug-in estimator of the extreme conditional quantile
in \cref{eq:vhat_A}.

Under Assumptions \ref{ass:compact} to 
\ref{ass:L1-id},
as $n \to \infty$, $k_n \to \infty$, 
$p_n= k_n/n \to 0$, and  $k_n/\sqrt{n} \to \infty$,
\[
      \hat\theta_n \;\xrightarrow{P}\; \theta_0,
\]
and for every fixed $q \in (0,1)$,
\[
\hat v_n(q\,|\,p_n) - v(q\,|\,p_n;C)
      \;\xrightarrow{P}\; 0.
\]
\ethm

\cref{thm:consistency_vqhat} treats the regime where $v(q \mid p;C)$ converges to zero. In this case, we establish consistency by showing the ratio converges to one in probability. Theorem 3.1 in 
\cite{nolde:2025} 
presents a similar result for the adjustment factor in the case $q = p$.

\bthm
\label{thm:consistency_vqhat}
Assume the model for tail dependence function $b(\cdot;\theta)$ is
correctly specified and assumed $\kappa=1$.

\begin{enumerate}[label=\textnormal{(A\arabic*)}]
 \item 
 The parameter space $\Theta\subset\mathbb R^{p}$ is a compact subset.
  \item There exists a true parameter $\theta_0\in\Theta \subset \mathbb{R}^p$ such that, for every  
        \[
            (w_1,w_2)\in\mathcal S_1
            \;:=\;
            \{(w_1,w_2):w_1>0,\;w_2>0,\;w_1+w_2=2\},
        \]
        \[
            \frac{C(u w_1,u w_2)}{u}
            \;=\;
            b(w_1,w_2;\theta_0)+R(u;w_1,w_2), 
            \qquad
            \sup_{(w_1,w_2)\in\mathcal S_1}\!|R(u;w_1,w_2)|
            \xrightarrow[u\downarrow 0]{}0 .
        \]
        \item The map $(w_1,w_2,\theta) \mapsto b(w_1,w_2;\theta)$ is continuous on  
        $\mathcal S_1\times\Theta$.
  \item   For every $\theta\in\Theta$ the function  
        $$H(r;\theta):=b(1,r;\theta),\;r>0,$$
        is a proper, absolutely continuous distribution function on $(0,\infty)$.
  \item  
  \label{ass:identifiability}
  For every $\varepsilon>0$,
$$
\inf _{\left\{\theta:\left\|\theta-\theta_0\right\| \geq \varepsilon\right\}} \int_{\mathcal{S}_1}\left|b\left(w_1, w_2 ; \theta\right)-b\left(w_1, w_2 ; \theta_0\right)\right| \mathrm{d} w_1 \mathrm{d} w_2>0.
$$
\end{enumerate}

Let $\hat{\theta}_n$ minimize the sample objective
$
  Q_{k_n,n}(\theta)
$
defined in \cref{eq:Q_alpha_n}, and let
$
  \hat v_n(q\mid p_n)
$
be the conditional quantile estimator in \cref{eq:v_q_hat},   
Then, as $n\to\infty$, $k_n\to\infty$, $k_n/n\to 0$,
       $$     \hat{\theta}_n\xrightarrow{P}\theta_0;
        $$
For every fixed $q\in(0,1)$, as $p_n \to 0$,
        \[
            \frac{\hat v_n(q\mid p_n)}{v(q\mid p_n)}\xrightarrow{P}1 .
   \]
\ethm
The finite-sample performance of the proposed estimation approach is assessed through an extensive simulation study. The parametric extreme value models, along with detailed estimation procedures and results, are reported in Section B of the supplementary material.

\section{Empirical systemic risk analysis} 
\label{sec:empirical}

This section applies the proposed framework to empirical data from the U.S. financial market. The goal is to assess systemic risk contributions and exposures using CoVaR and $\Delta$CoVaR, with an emphasis on how joint tail behavior and tail dependence with the system shape these measures and reveal differences in the systemic roles of assets and institutions.


\subsection{CoVaR and $\Delta$CoVaR in the Copula-AR-GARCH model}

Let $\mathbf{R}_t = (R_{ti}, R_{tS})$ denote the bivariate return vector at time $t$, consisting of returns for institution $i$ and the system $S$, for $t = 1, 2, \ldots, T$. The return process is adapted to the filtration $\{\mathcal{F}_t\}_{t=0}^T$, generated by the observed history:
$
\mathcal{F}_t := \sigma\left( \mathbf{R}_s : s \le t \right).
$
Under the Copula-AR-GARCH model \citep{jondeau2006copula}:
\begin{align}
R_{tj} &= \mu_{tj} + \sigma_{tj} Z_{tj}, \qquad Z_{tj} \sim F_{Z_j}, \quad j \in \{i, S\},
\end{align}
where: $\mu_{tj} = \EE(R_{tj} \mid \mathcal{F}_{t-1})$,  $\sigma_{tj}^2 = \operatorname{Var}(R_{tj} \mid \mathcal{F}_{t-1})$, and $Z_{tj}$ are standardized innovations with marginal distribution $F_{Z_j}$.
The AR and GARCH dynamics are given by:
$$
\mu_{tj} = \mu_j + \sum_{k=1}^{p_j} \phi_{jk} R_{t-k,j}, \quad
\sigma_{tj}^2 = \beta_{0j} + \beta_{1j} Z_{t-1,j}^2 + \beta_{2j} \sigma_{t-1,j}^2.
$$

The joint conditional distribution of returns $\left(R_{t i}, R_{t S}\right)$ is determined by the copula $C_{i S}$ linking the innovations:
$$
F_{\mathbf{R}_t}\left(r_{t i}, r_{t S} \mid \mathcal{F}_{t-1}\right)=C_{i S}\Big(F_{Z_i}\big(\frac{r_{t i}-\mu_{t i}}{\sigma_{t i}}\big), F_{Z_S}\big(\frac{r_{t S}-\mu_{t S}}{\sigma_{t S}}\big)\Big) .
$$

Let the distress event of institution $i$, given the past information, be
$$
D_{i, t}(p):=\left\{R_{ ti} \leq  \operatorname{VaR}_{i \mid t}(p)\right\}
$$
and the enlarged information set be the smallest $\sigma$-field that includes both the past market information and the event that institution $i$ is in distress:
$$
\mathcal{H}_{i,t}(p):=\mathcal{F}_{t-1} \vee \sigma\left(D_{i, t}(p)\right)=\sigma\left(\mathcal{F}_{t-1} \cup\left\{D_{i, t}(p)\right\}\right) .
$$
Then, the system CoVaR conditional on the information set $\mathcal{H}_{i, t}(p)$ is 
\begin{align}
 \operatorname{CoVaR}_{S \mid i, t}(p)=\mu_{t S}+\sigma_{t S} F_{Z_S}^{-1}\big(r_{S \mid i}(p) \cdot p\big)
 \label{eq:covar_garch}
\end{align}
with the adjustment factor 
$
r_{S \mid  i}(p)=C_{S \mid i \leq}^{\quad \gets}(p \mid p)/p.
$

The time-varying $\Delta$CoVaR is given by 
\begin{align}
    \Delta \covar_{S \mid i, t}(p)=
    \frac{\covar_{S \mid i, t}(p)-\var_{S \mid t}(p)}{|\var_{S \mid t}(p)|}=\frac{
    \sigma_{t S} F_{Z_S}^{-1}(r_{S \mid i}(p) \cdot p)- \sigma_{t S} F_{Z_S}^{-1}(p)
    }{|\mu_{t S}+\sigma_{t S} F_{Z_S}^{-1}(p)|}.
\end{align}
For financial returns, the conditional mean 
$\mu_{tS}$ is typically negligible relative to the tail quantiles, so 
we obtain,
\begin{align}
\Delta \operatorname{CoVaR}_{S \mid i}(p) \approx \frac{F_{Z_S}^{-1}\left(r_{S \mid i}(p) \cdot p\right)-F_{Z_S}^{-1}(p)}{|F_{Z_S}^{-1}(p)|},
\label{eq:deltacovar_garch}
\end{align}
and \cref{thm:deltaCoVaR} would apply.

The explicit formulas clarify the main drivers of the two risk measures and how they should be interpreted.
$\covar_{S \mid i, t}(p)$ in \cref{eq:covar_garch} measures the absolute level of system loss conditional on institution $i$ being in distress. As a high-frequency risk metric, it is largely driven by changes in market volatility.
$\Delta\operatorname{CoVaR}_{S \mid i}(p)$ in \cref{eq:deltacovar_garch}, on the other hand,
is determined more by the dependence and market regime,
captured by the adjustment factor $r_{S \mid i}(p)$ and the tail index of $F_{Z_S}$.
A smaller adjustment factor and a larger tail index both contribute to a more negative $\Delta\covar_{S \mid i}$,  
indicating a greater contribution of firm-level distress to systemic risk.


\begin{figure}
    \centering
    \begin{subfigure}[t]{0.45\linewidth}
\centering        \includegraphics[width=\linewidth]{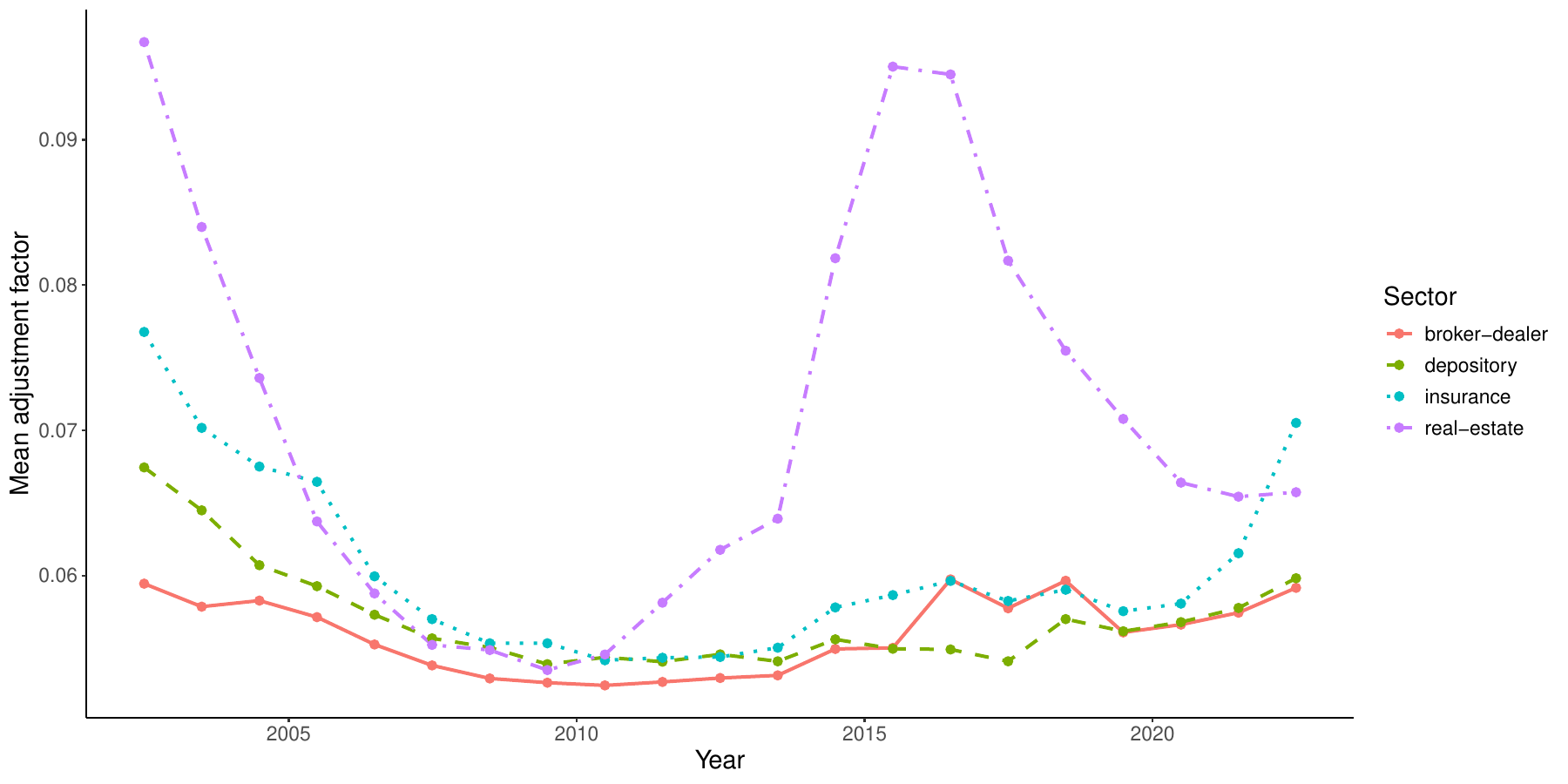}
    \end{subfigure}
    \hfill
    \begin{subfigure}[t]{0.45\linewidth}
\centering       \includegraphics[width=\linewidth]{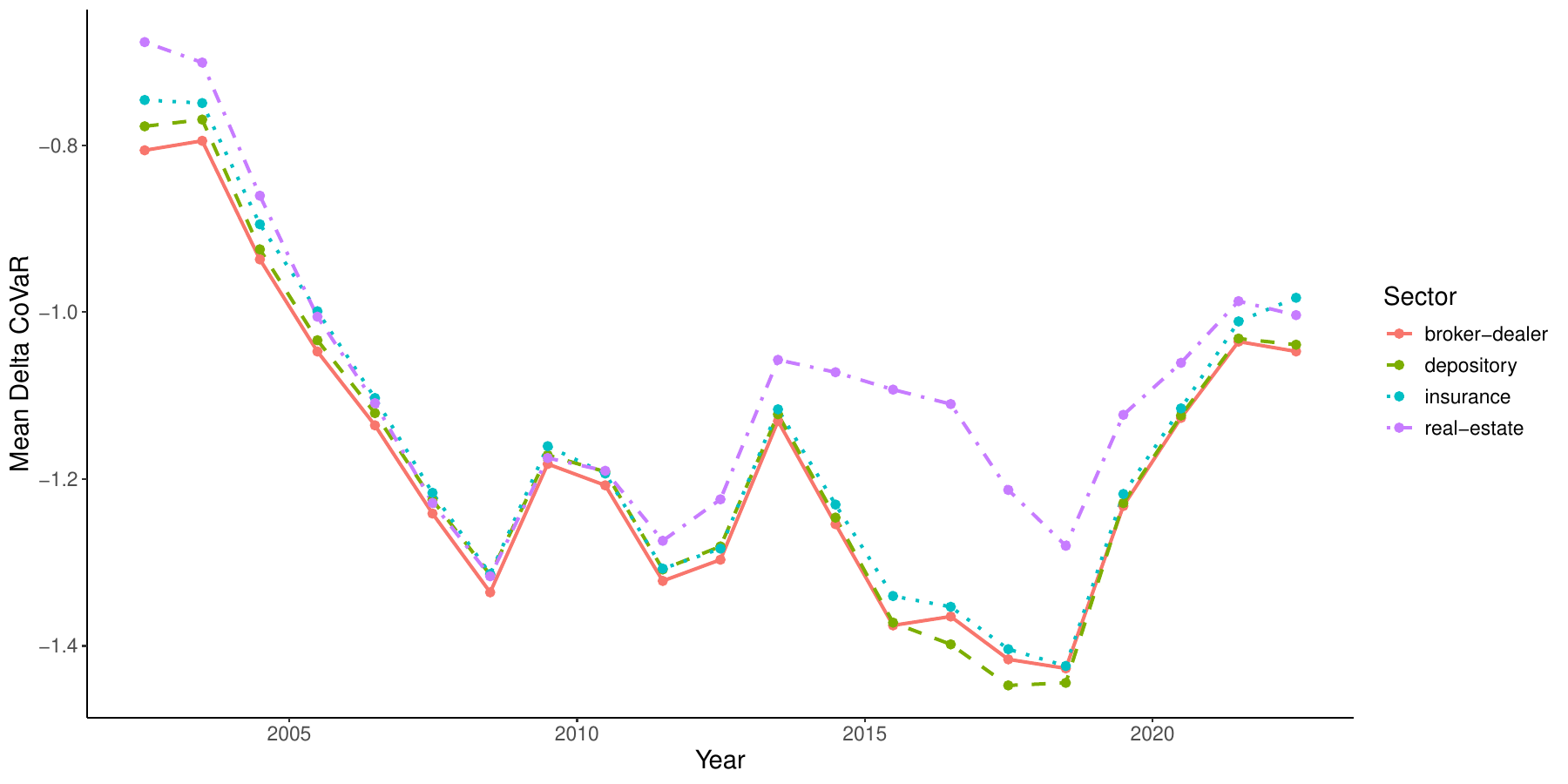}
    \end{subfigure}
  \caption{Mean estimates of 
  $r_{S\mid i}(p)$ (left) and 
  $\Delta \covar_{S \mid i}(p)$(right) for large U.S. financial institutions
   with $p=0.05$, segmented by depository, broker-dealers, insurance companies, and real estate firms. 
The estimates for $r_{S\mid i}(p)$ and  
 $\Delta \covar_{S \mid i}(p)$ are calculated based on the daily log return data from June 2000 to June 2025 using five-year rolling window.}
   \label{fig:covar_finance_sector}
\end{figure}

\begin{figure}
    \centering    \includegraphics[width=0.7\linewidth]{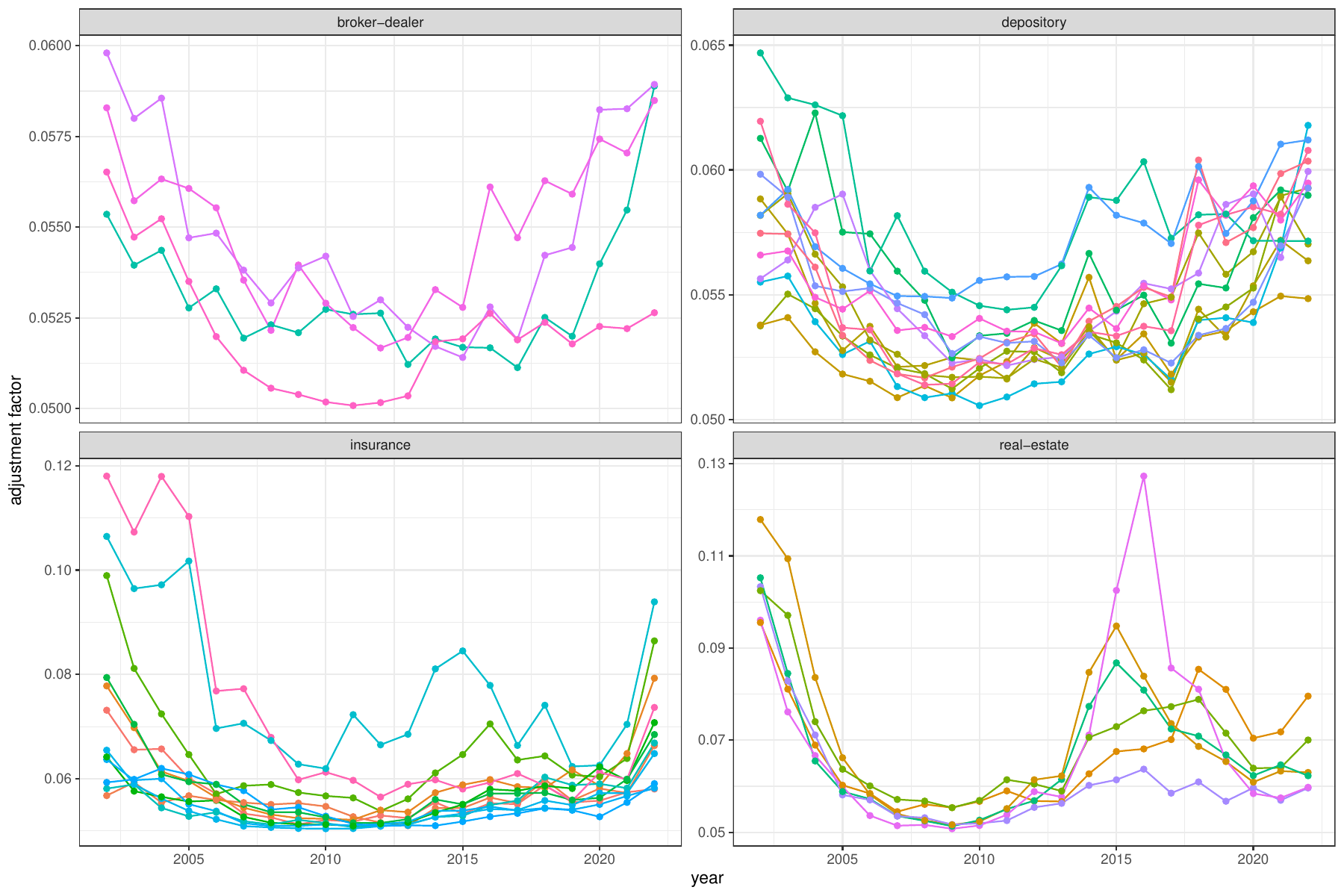}
   \caption{
   Path of the estimated adjustment factor $r_{S \mid i}(p)$ at $p=0.05$ for 35 institutions that are consistently among the top 100 firms by market capitalization throughout 2000-2025.
}\label{fig:rp_path_finance_sector}
\end{figure}

\subsection{Systemic risk impact of the financial sector}
\label{sec:6-2}

A large literature, e.g., \cite{bernal2014assessing}, documents the buildup of systemic risk in the financial sector before the 2008 crisis, and the sector remains central to macroprudential regulation. However, structural changes in the economy and the rapid expansion of the technology sector raise the question of whether the systemic role of financial institutions has shifted over time. 
We address this question by estimating $\Delta \covar_{S \mid i}$ for large U.S. financial institutions, using the S\&P 500 index as a proxy for the system. For each rolling five-year window over 2000--2025, we identify the 100 largest firms by market capitalization and model their daily returns using marginal AR-GARCH specifications with skew- $t$ innovations. Based on the resulting filtered innovations, we then estimate the adjustment factor $r_{S \mid i}(p)$ following the procedure proposed in  \cref{sec:5}.

Candidate models are derived from the tail expansions of parametric copulas listed in \cref{tab:copula_expansions}, with further details provided in Section B.1 of the Supplementary Material. As discussed in \cref{sec:5-2}, the model choices are determined by the tail regime. 
In practice, we tell the regime based on the pair of empirical tail dependence coefficients in the lower-left and upper-left corners, ($\hat{\lambda}, \hat{\lambda}^{2 *}$). A positive $\hat{\lambda}$ with $\hat{\lambda}^{2 *}$ close to zero indicates tail attraction, in which case $b\left(w_1, w_2\right)$ is modeled using the 
lower tail dependence function of the 
reflected Gumbel or Student-$t$ copula. A positive $\hat{\lambda}^{2 *}$ with $\hat{\lambda}$ close to zero indicates tail repulsion, and $b^{* 2}\left(w_1, w_2\right)$ is modeled using the lower tail dependence function of the Clayton copula. If both coefficients are close to zero, the regime is classified as tail balance, and $A(v)$ is modeled using the boundary cdf of the reflected IPS copula. 
Following the simulation experiments in the Supplementary Material, $k_n=100$ is used in the analysis.

\cref{fig:covar_finance_sector} 
reports the mean estimates of $r_{S \mid i}(p)$ and $\Delta \covar_{S \mid i}(p)$ at $p=0.05$, grouped by Standard Industrial Classification (SIC) into depository institutions, broker-dealers, insurance firms, and real estate firms. \cref{fig:rp_path_finance_sector}
as complementary to the aggregate view, shows the time path of $r_{S \mid i}(p)$ for institutions that remained in the top 100 by market capitalization throughout 2000--2025.

Overall, it suggests that the financial sector remains the major contributor to systemic risk with the adjustment factors close to the lower bound $p=0.05$ (compare the comonotonicity bound in \cref{prop:properties-of-v(q|p)}
when $q=p$)
and large negative values of $\Delta \mathrm{CoVaR}_{S \mid i}$. Over time, as shown in \cref{fig:rp_path_finance_sector},
both banks and insurance firms exhibit a mild U-shaped pattern: their systemic importance rises from the early 2000s, peaks around the financial crisis and its aftermath, and then declines gradually in recent years. Real-estate firms, however, show a different pattern. Their systemic impact increases sharply in the years leading up to the crisis, declines markedly afterward, and then remains relatively stable in recent years.

\subsection{Cross-sectional contributions to systemic risk}

\begin{figure}[!t]
    \centering
    \begin{subfigure}[t]{0.45\linewidth}
        \centering
\includegraphics[width=\linewidth]{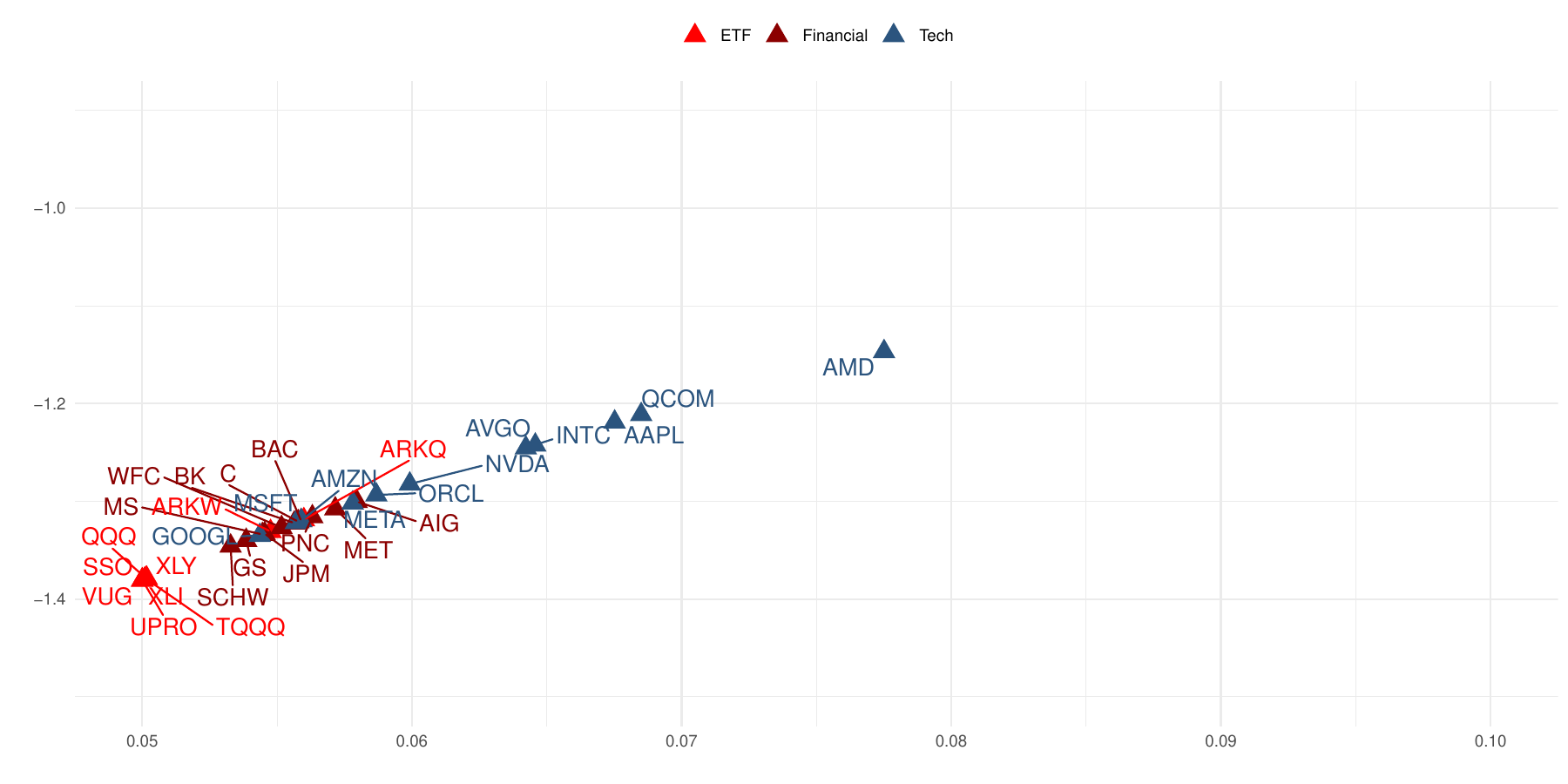}
    \end{subfigure}
    \hfill
    \begin{subfigure}[t]{0.45\linewidth}
        \centering
\includegraphics[width=\linewidth]{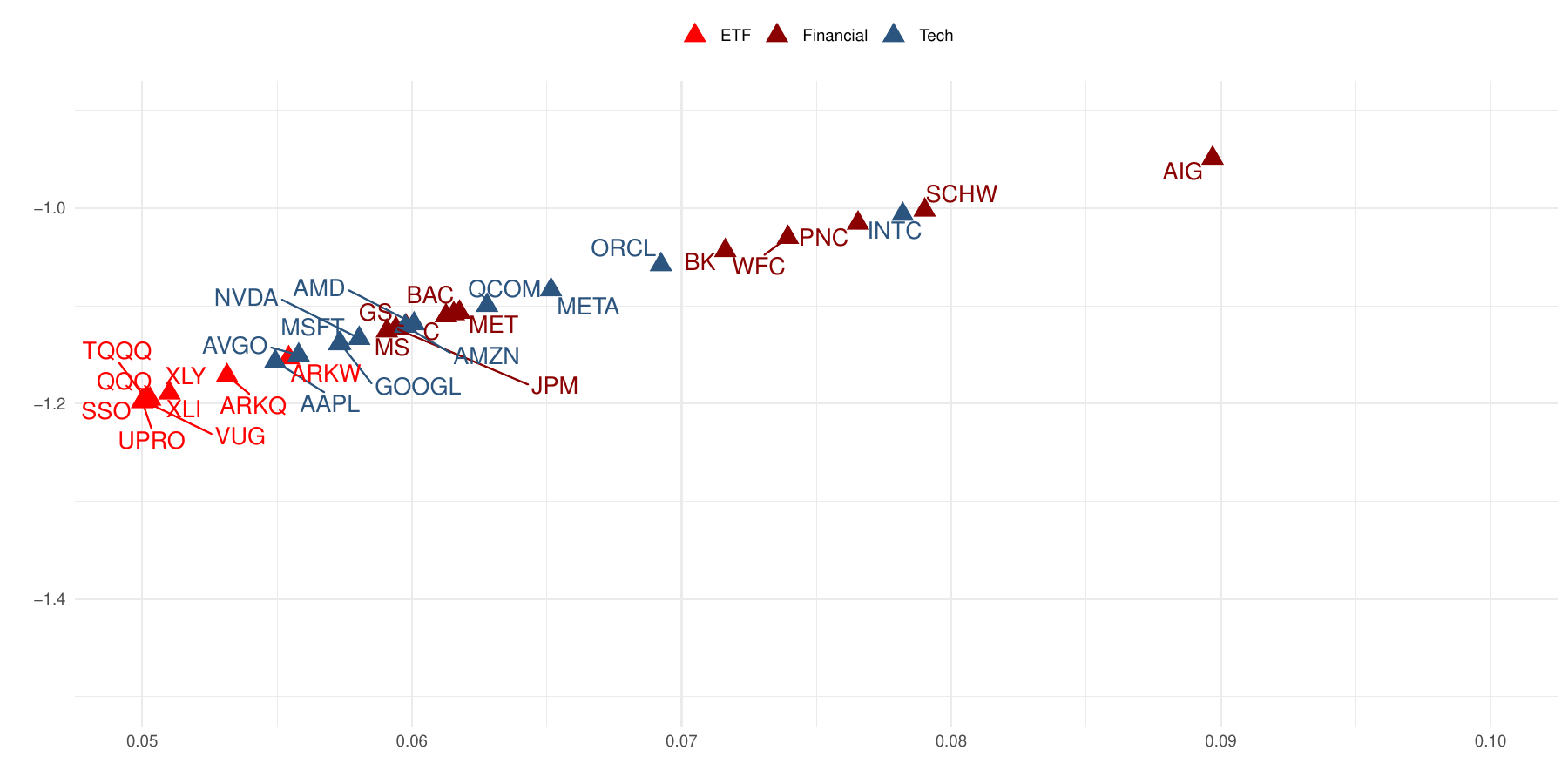}
    \end{subfigure}
    \caption{$\operatorname{CoVaR}_{S \mid i}(p)$ versus the adjustment factor $r_{S \mid i}(p)$ at $p = 0.05$ for major U.S. financial institutions, large technology firms, and equity ETFs,  based on daily returns for two sample periods: 2012–2018 (left panel) and 2018–2025 (right panel).
}
    \label{fig:covar_impact_1}
\end{figure}

\begin{figure}
    \centering
    \begin{subfigure}[t]{0.45\linewidth}
\centering \includegraphics[width=\linewidth]{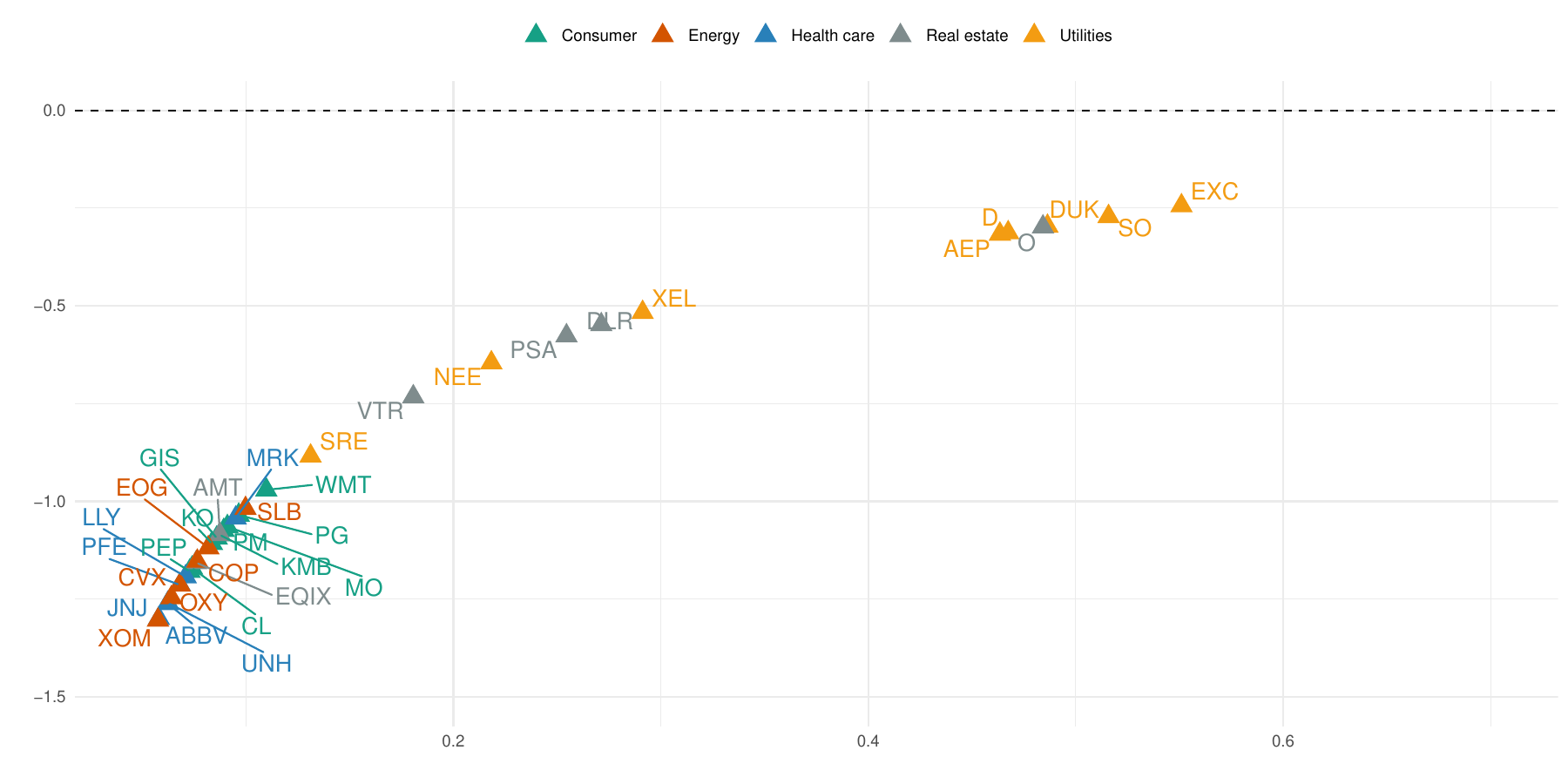}
    \end{subfigure}
    \hfill
    \begin{subfigure}[t]{0.45\linewidth}
\centering \includegraphics[width=\linewidth]{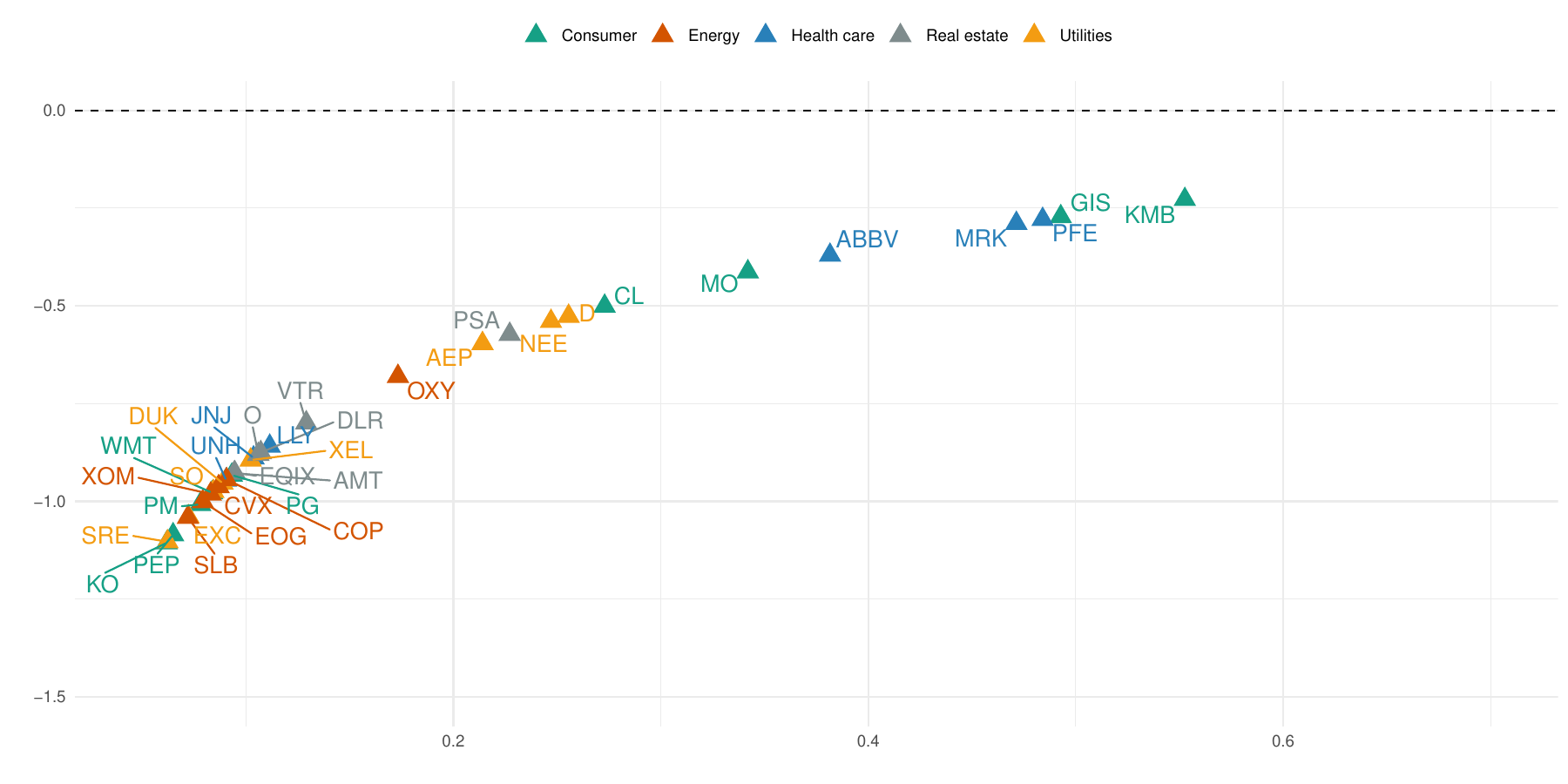}
    \end{subfigure}
    \caption{$\Delta \operatorname{CoVaR}_{S \mid i}(p)$ versus the adjustment factor $r_{S \mid i}(p)$ at $p=0.05$ for firms across consumer, energy, health care, real estate, and utilities sectors, based on daily returns for two sample periods: 2012–2018 (left panel) and 2018–2025 (right panel). }
    \label{fig:covar_impact_2}
\end{figure}

This section reveals cross-sectional differences in systemic risk contributions among large-capitalization U.S. firms across major sectors.
\cref{fig:covar_impact_1} and \cref{fig:covar_impact_2} plot the estimated $\Delta \mathrm{CoVaR}_{S\mid i}(p)$ against the adjustment factor $r_{S\mid i}(p)$ at $p=0.05$ for firms spanning financials, technology, consumer, energy, health care, real estate, and utilities over two sample periods, 2012–2018 and 2018–2025.

As in \cref{fig:covar_impact_1},  financial institutions and technology firms exhibit the strongest tail dependence with the system, and this translates to strong tail attraction between firm-level distress and system losses.
It signals that these firms are likely systemically central: stress originating in them tends to transmit rapidly and intensify system-wide tail losses, making them potential originators and amplifiers of systemic risk.

Firms from the other sectors, as shown in \cref{fig:covar_impact_2}, also exhibit negative $\Delta \mathrm{CoVaR}_{S \mid i}$, indicating a positive contribution to systemic risk. But their tail behaviors are more heterogeneous, combining both tail attraction and tail balance, and their adjustment factors and $\Delta \mathrm{CoVaR}_{S \mid i}$ span a wider range.

In \cref{fig:covar_impact_1}, we notice that the entities with the most negative $\Delta \mathrm{CoVaR}_{S \mid i}$ are large equity exchange-traded funds (ETFs) and mutual funds.
As discussed in financial literature \citep{ramaswamy2011market}, these investment vehicles typically do not originate distress, but they act as important channels that amplify systemic losses.

Comparing the two periods, $\Delta \mathrm{CoVaR}_{S \mid i}$ exhibits an overall upward shift. This change is because the innovation distribution has a less heavier tail in 2018-2025 compared to 2012-2018. The behavior of $r_{S \mid i}(p)$ reveals changes in the dependence structure. In particular
we observe a clear reordering of systemic importance. In 2012--2018, systemic risk contributions are dominated by financial institutions, while most technology firms, aside from a few exceptionally large firms such as Microsoft and Google, have relatively modest impacts. This pattern changes markedly in 2018--2025, with systemic importance shifting toward the technology sector and away from financial firms. As shown in \cref{fig:covar_impact_2}, this trend is particularly evident for companies such as NVIDIA, AMD, and Broadcom, which have expanded rapidly with surging demand for AI infrastructure.

This shift is consistent with the findings in \cref{sec:6-2}, which indicate a declining systemic impact of the financial sector. At the same time, it points to an emerging source of systemic risk in the technology sector, particularly among firms closely tied to the AI boom.

\subsection{Systemic risk exposures and hedging assets}

The $\Delta \mathrm{CoVaR}$ in \cref{eq:deltacovar_garch}, when conditioning on system-wide distress, measures an asset’s exposure to systemic risk. A value near zero indicates that the asset's lower-tail distribution is largely unaffected by systemic stress while a large positive $\Delta \mathrm{CoVaR}_{i \mid S}$ implies that the asset tends to gain value when the system is under distress.

\begin{figure}[!t]
    \centering
    \begin{subfigure}[t]{0.45\linewidth}
        \centering     \includegraphics[width=\linewidth]{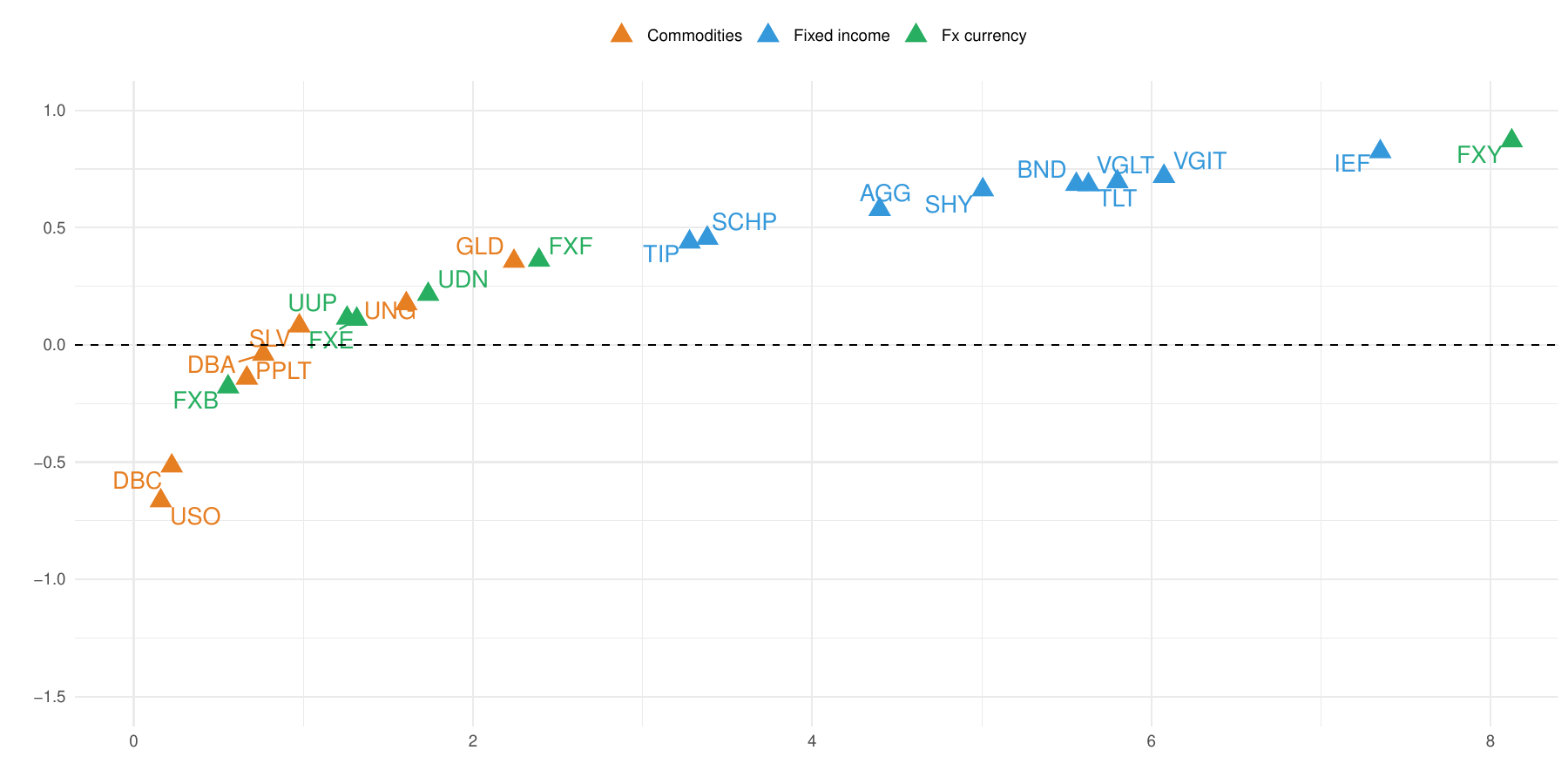}
    \end{subfigure}
    \hfill
    \begin{subfigure}[t]{0.45\linewidth}
        \centering       \includegraphics[width=\linewidth]{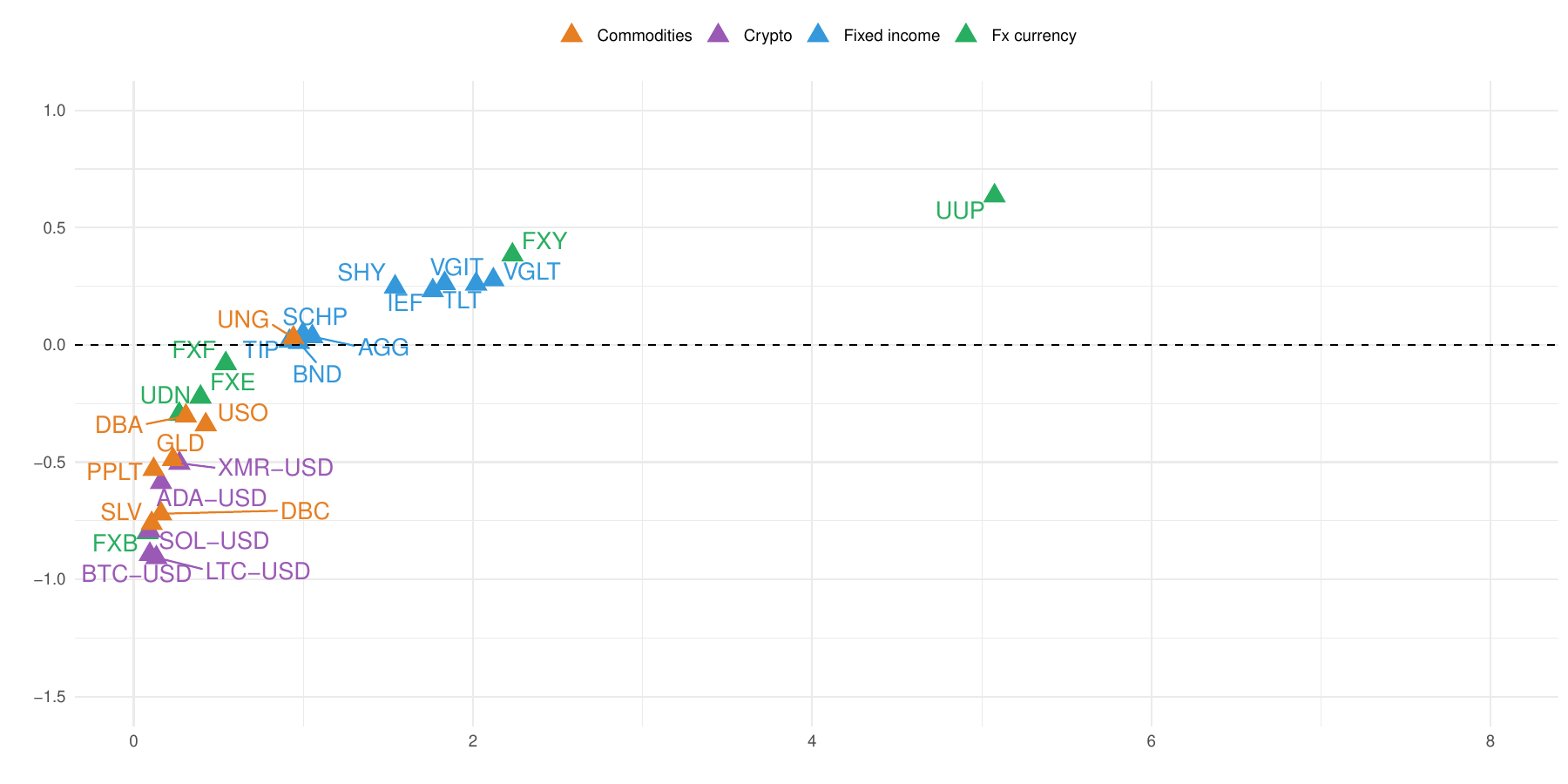}
    \end{subfigure}
    \caption{$\Delta \operatorname{CoVaR}_{i \mid S}(p)$ versus the adjustment factor $r_{i \mid S}(p)$ at $p=0.05$ for fixed income securities, cryptocurrency, commodities, computed based the daily returns from 2012--2018 (left panel) and 2018--2025 (right panel).}
    \label{fig:covar_exposure_1}
\end{figure}

\begin{figure}
    \centering
    \begin{subfigure}[t]{0.45\linewidth}
        \centering       \includegraphics[width=\linewidth]{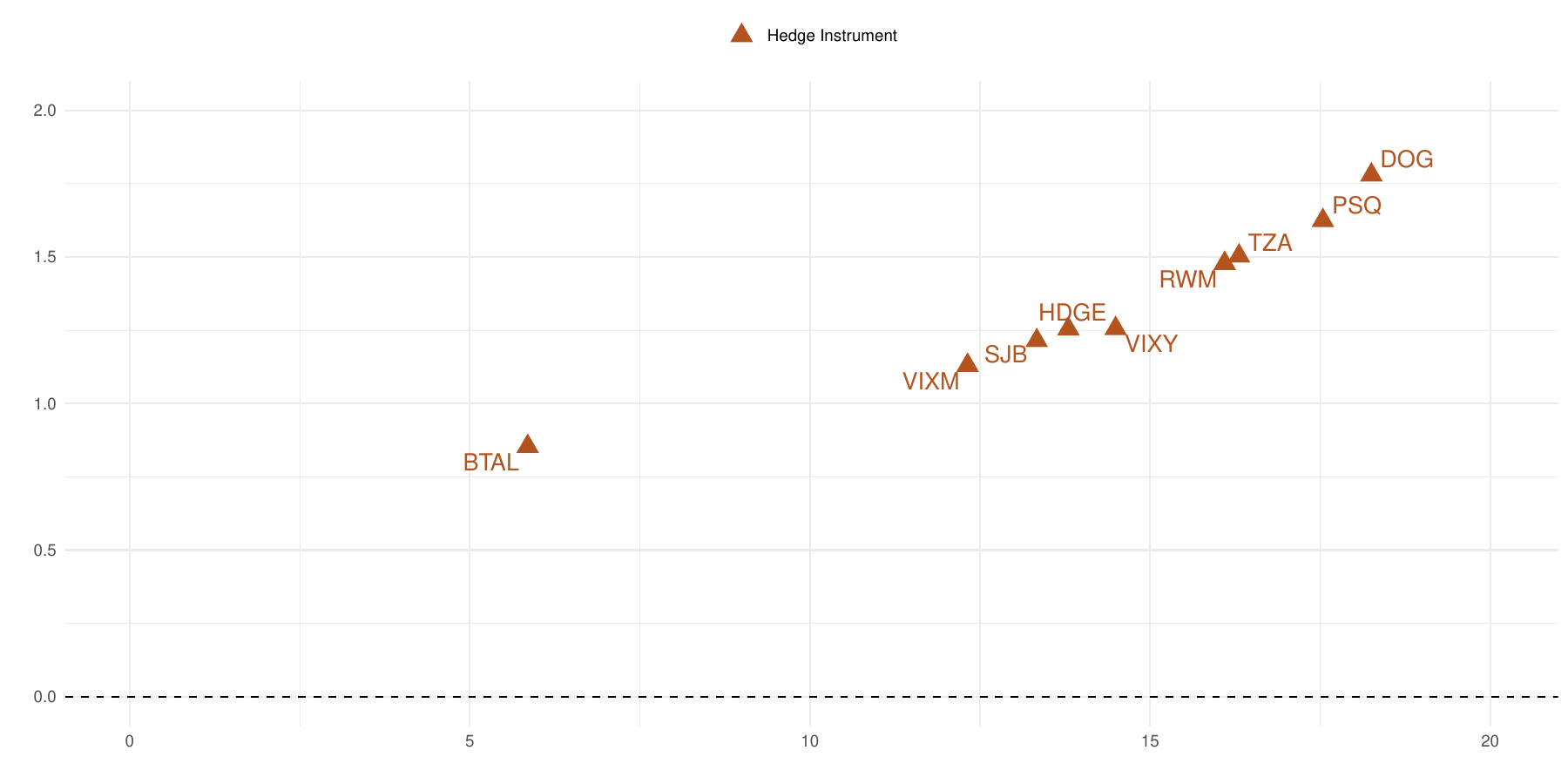}
    \end{subfigure}
    \hfill
    \begin{subfigure}[t]{0.45\linewidth}
        \centering     \includegraphics[width=\linewidth]{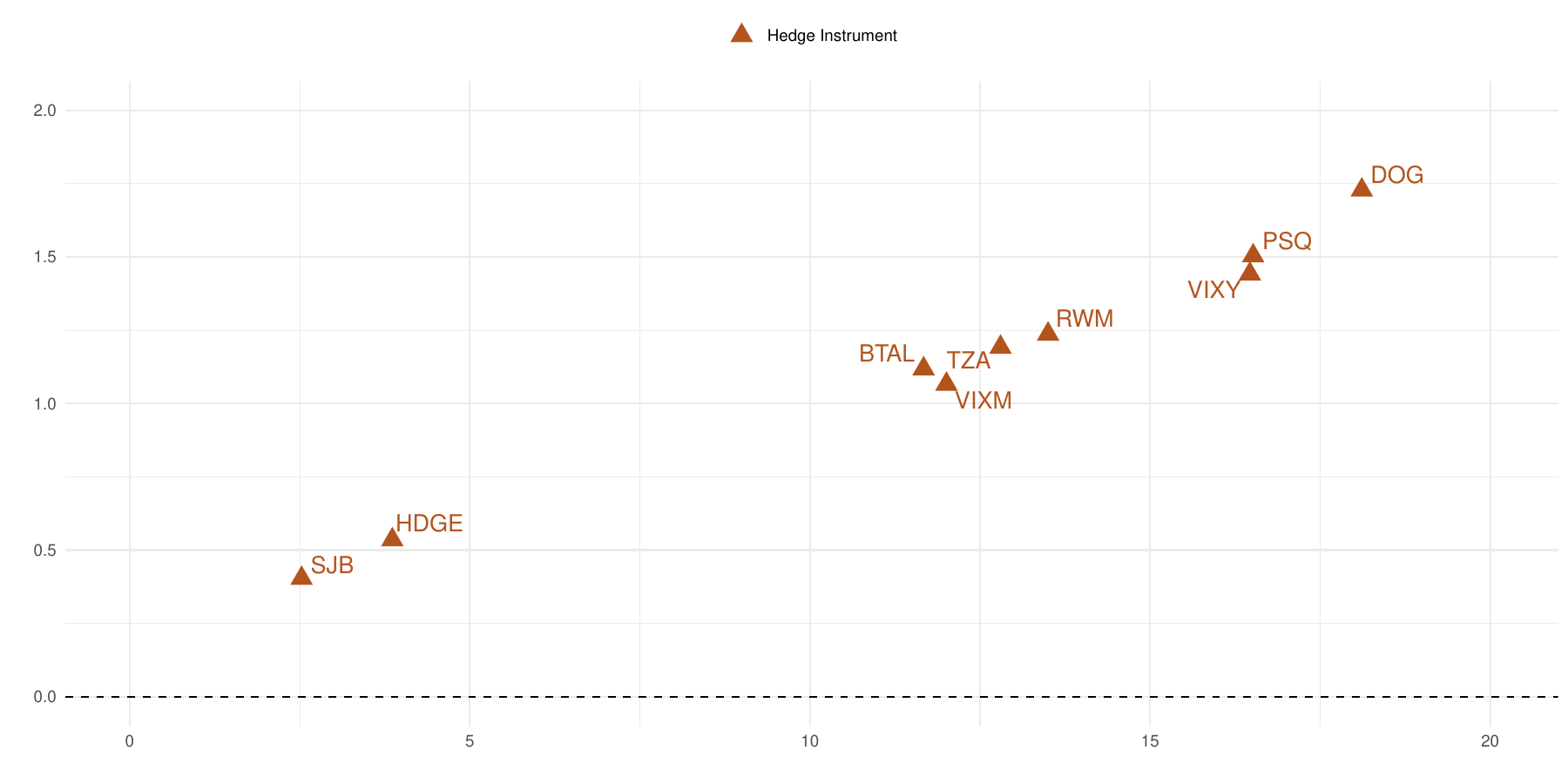}
    \end{subfigure}
     \caption{$\Delta \operatorname{CoVaR}_{i \mid S}(p)$ versus the adjustment factor $r_{i \mid S}(p)$ at $p=0.05$ for hedging instruments based on daily returns from 2012--2018 and 2018--2025.}      \label{fig:covar_exposure_2}
\end{figure}

This information is useful for portfolio risk management as
 $\mathrm{CoVaR}_{i \mid S}$ can help investors and risk managers identify assets that improve portfolio resilience during episodes of system-wide stress. \cref{fig:covar_exposure_1} and \cref{fig:covar_exposure_2}
plot the estimated $\Delta \mathrm{CoVaR}_{i \mid S}(p)$ against $r_{i \mid S}(p)$ at $p=0.05$, for a wide range of assets including fixed income securities, commodities, and cryptocurrencies based on daily returns over 2012--2018 and 2018--2025.


For structured hedging instruments shown in \cref{fig:covar_exposure_2}, we observe strong tail repulsion and large positive $\Delta \mathrm{CoVaR}_{i \mid S}$. Some of these instruments are explicitly designed to move against the system, such as inverse equity ETFs and VIX‑linked products. Others provide protection through exposure to specific stress channels; for example, SJB, which holds short positions in high‑yield corporate bonds, gains from the deterioration of credit markets during systemic crises.

\cref{fig:covar_exposure_1} identifies several natural hedging assets, including long-duration U.S. Treasury bond ETFs (e.g., TLT, VGLT, and IEF) and currency ETFs (e.g., FXY, FXF). These assets exhibit weak tail repulsion, or negative tail balance, indicating that they tend to appreciate or remain stable under systemic distress.

The right panel of \cref{fig:covar_exposure_1} reports the estimates for cryptocurrencies. Although these assets are often viewed as isolated from the traditional financial assets \citep{corbet2018exploring}, our results show that they still display weak tail attraction, or positive tail balance. In addition, their marginal tail indices are noticeably larger than those of other asset classes, pointing to greater exposure to idiosyncratic extreme shocks. 

The sign reversal in the bond-stock correlation has been widely documented in the economics literature, see e.g., \cite{campbell2025bond},
and is often linked to shifts in the macroeconomic regime. Comparing the two panels of \cref{fig:covar_exposure_1} suggest a similar change in tail dependence. The $r_{i \mid S}(p)$ of fixed-income bonds and safe-haven currencies 
decreases and $\Delta \mathrm{CoVaR}_{i \mid S}$
becomes less positive, and in some cases, turns negative. This indicates that their hedging effectiveness against systemic risk declines over time, with these assets increasingly comoving with system losses instead of offsetting them.

\section{Discussion}
\label{sec:discussion}

In this work, we develop a theoretical and inferential framework for conditional Value-at-Risk (CoVaR) based on copula and extreme value theory.
It clarifies how the tail dependence structure determines the limiting behavior of CoVaR and $\Delta$CoVaR, thereby providing a principled basis for interpreting these measures in systemic risk analysis.

The paper also contributes an empirical analysis based on the proposed framework, highlighting several important findings. Our results show that, although the financial sector remains the major contributor to systemic risk, its impact has weakened in recent years, coinciding with the rising systemic importance of major technology firms.
From the perspective of macroprudential regulation, this shift suggests an emerging source of systemic risk from the technology sector.
The 2007–2009 financial crisis illustrates how the repricing of housing assets became systemic when compounded with the structure of the financial system.
Our findings point to a related but distinct concern: even though the shocks originate outside the traditional financial sector, they may generate system-wide effects as major technology firms have become deeply embedded in the system’s downside dependence structure. An important direction for future research is therefore to develop systemic risk monitoring and stress-testing frameworks that account for the potential repricing of AI-linked assets and their impacts for system-wide stability.

\subsection*{Data availability statement}
The data used in this study were obtained through Wharton Research Data Services (WRDS) and Yahoo Finance.
The replication package, including the code and instructions for accessing the data, is available in a GitHub repository.
\subsection*{Disclosure statement}
The authors report there are no competing interests to declare.



\bibliography{bibliography.bib}

\end{document}